\documentclass[reprint,amssymb,aps]{revtex4-1}

\pdfoutput=1

\usepackage{amsmath}
\usepackage{bm}
\usepackage{multirow}
\usepackage{bigdelim}
\usepackage{enumitem}

\newcommand{\overbar}[1]{\mkern 1.5mu\overline{\mkern-1.5mu#1\mkern-1.5mu}\mkern 1.5mu}

\begin{document}

\title{
  New quantum number for the many-electron Dirac-Coulomb Hamiltonian
}

\author{Stanislav Komorovsky}
  \email{stanislav.komorovsky@savba.sk}
  \affiliation{
    Centre for Theoretical and Computational Chemistry, Department of Chemistry,
    UiT The Arctic University of Norway, N-9037 Troms{\o}, Norway
  }
  \affiliation{
    Institute of Inorganic Chemistry, Slovak Academy of Sciences, D\'ubravsk\'a
    cesta 9, SK-84536 Bratislava, Slovakia
  }
\author{Michal Repisky}
  \affiliation{
    Centre for Theoretical and Computational Chemistry, Department of Chemistry,
    UiT The Arctic University of Norway, N-9037 Troms{\o}, Norway
  }

\author{Luk\'a\v{s} Bu\v{c}insk\'y}
  \affiliation{
    Institute of Physical Chemistry and Chemical Physics, Slovak University of
    Technology, Radlinskeho 9, Bratislava, SK-812 37, Slovakia
  }

\date{\today}

\begin{abstract}

By breaking the spin symmetry in the relativistic domain, a powerful tool in
physical sciences was lost. In this work, we examine an alternative of spin
symmetry for systems described by the many-electron Dirac-Coulomb
Hamiltonian.  We show that the square of many-electron operator
$\mathcal{K}_+$, defined as a \emph{sum} of individual single-electron time-reversal
(TR) operators, is a linear Hermitian operator which commutes with the
Dirac-Coulomb Hamiltonian in a finite Fock subspace.  In contrast to the
square of a standard unitary many-electron TR operator $\mathcal{K}$, the
$\mathcal{K}^2_+$ has a rich eigenspectrum having potential to substitute spin
symmetry in the relativistic domain.  We demonstrate that $\mathcal{K}_+$ is
connected to $\mathcal{K}$ through an exponential mapping, in the same way
as spin operators are mapped to the spin rotational group. Consequently, we
call $\mathcal{K}_+$ the generator of the many-electron TR symmetry.  By
diagonalizing the operator $\mathcal{K}^2_+$ in the basis of Kramers-restricted
Slater determinants, we introduce the relativistic variant of configuration
state functions (CSF), denoted as Kramers CSF.  A new quantum number associated
with $\mathcal{K}^2_+$ has potential to be used in many areas, for instance, (a)
to design effective spin Hamiltonians for electron spin resonance spectroscopy
of heavy-element containing systems; (b) to increase efficiency of methods for
the solution of many-electron problems in relativistic computational chemistry
and physics; (c) to define Kramers contamination in unrestricted
density functional and Hartree--Fock theory as a relativistic analog of the
spin contamination in the nonrelativistic domain.

\end{abstract}


\maketitle


\section{\label{sec:intro}Introduction}

In experimental and theoretical science, it is of great importance to know the
symmetry of the system studied. In spectroscopy, the choice of the effective
Hamiltonian, used to fit the experimental data, is influenced (if not based) by
the symmetry of the system. Similarly, taking symmetry into account in quantum
computational science increases the efficiency and stability of computational
methods.

In this work, our attention points towards symmetries of electronic
systems in the absence of magnetic fields, while excluding symmetries
associated with external electric fields such as point-group symmetry or
translational symmetry given by clamped nuclei. To this class of symmetries we
consider, for example, spin symmetry in nonrelativistic and time-reversal
symmetry~\cite{Kramers1930, Wigner1932} in both the nonrelativistic and the
Dirac four-component relativistic level of theory~\cite{DyallFaegri2007,
Schwerdtfeger2002, ReiherWolf2009}.  Here, the following general rule applies:
all symmetries present at the higher level of theory appear also at lower level
of theory or, in other words, going from a lower to a higher level of theory
can lead to symmetry breaking.

Ideally, every study of a quantum system should use the highest possible level of
theory. However, in practical applications, it is common to restrict the type of
Hamiltonian and the representation of the wave function to make a given
calculation feasible. The decision factor usually is the energy scale
in combination with the accuracy needed for the problem under investigation.
This work is aimed at the domain of relativistic quantum chemistry, and
therefore we only consider theories which include spin-orbit (SO) interaction
non-perturbatively (two- or four-component) and partly theories without SO
interaction (nonrelativistic or scalar relativistic). A quantum
electrodynamics theory and other particle theories are beyond the scope of this
study.

The nonrelativistic electronic structure theory with spin introduced {\it ad
hoc}, has been thoroughly investigated with respect to electron correlation,
system dynamics, spectroscopic parameters, and the theory of spin itself
has been worked out in great detail~\cite{Pauncz1979, Yamanouchi1936,
Lowdin1955a, *Lowdin1955b, Young1931, Wigner1959, Serber1934, Salmon1972}.
Eigenfunctions and eigenvalues of spin operators are well-known and are
successfully accommodated in different
spectroscopies~\cite{McHale1999,EPRbook,EPR_Bleaney}. Nevertheless, spin
symmetry is not appropriate for treating problems where SO effects become
non-negligible since, in this case, spin is no longer a good quantum number.
Although the time-reversal operator commutes with Hamiltonians accounting
explicitly for SO coupling, it is an antilinear operator and therefore does
not have eigenvalues and eigenvectors. For this reason, time-reversal symmetry
has never played as important a role in atomic and molecular spectroscopy as spin
symmetry. The same conclusions hold for the well-known generalization of the
time-reversal operator to the many-electron case (constructed as a \emph{product}
of the one-electron time-reversal operators,
$\mathcal{K}$)~\cite{Sachs1987,Lax2001}. Although the square of $\mathcal{K}$
becomes linear Hermitian operator and still commutes with the relativistic
many-electron Hamiltonian, the quantum number associated with the operator
contains very little information, as it is either $+1$ or $-1$ depending on the
even or odd number of electrons in the system~\cite{Sachs1987, Lax2001}.
Therefore, even the $\mathcal{K}^2$ can not substitute the role of the spin
operators in relativistic theories.

Still, time-reversal (TR) symmetry has been shown to simplify the evaluation
of matrix elements \cite{Rosch1980, Malvetti1989, DyallNATO,
Visscher1995, Aucar1995, Visscher1996, Jensen1996, Kim1998, Saue1999,
Fleig2001, Kim2003, Peng2009} and has been worked out in combination with
double-group symmetry \cite{Onodera1966, OregI, *OregII, *OregIII, Rosen1978,
Pitzer1988, DyallFaegri2007}. Nevertheless, the lack of useful quantum numbers
for the many-fermion open-shell wave function in the framework of relativistic
theories that account for spin-orbit coupling remains an obstacle. In the works
of Bu{\v c}insk{\' y} {\it et al.} \cite{Bucinsky2015, Bucinsky2016}, a new
operator $\mathcal{K}_+$, suitable for treating open-shell systems, has been
proposed. This operator has originally been denoted as a {\it pseudo-time-reversal
operator}. However, in this work we use the name {\it
time-reversal generator} to emphasize the fact that it generates the TR
operator $\mathcal{K}$ in similar way as spin operators generate operators of
spin rotations.  The operator $\mathcal{K}_+$ is constructed as a \emph{sum} of
individual one-electron time-reversal operators and its square produces quantum
number with information which supersedes the well-known counterpart
$\mathcal{K}^2$.  Eigenfunctions (in the basis of Kramers-restricted Slater
determinants) and eigenvalues of the $\mathcal{K}^2_+$ operator have been
presented previously for cases with up to four unpaired electrons with all
possible Kramer determinants. However, these eigenfunctions were built
phenomenologically and were not orthonormalized.  Herein, we give additional
insight into the relation between the many-electron $\mathcal{K}$ and
$\mathcal{K}_+$ operators. Furthermore, we investigate the spectrum of the
square of the time-reversal generator $\mathcal{K}^2_+$ both for general wave
functions and for the basis of Kramers-restricted determinants. We show the
commutation relation between the Dirac-Coulomb Hamiltonian and the
$\mathcal{K}^2_+$ operator, thus introducing a new quantum number associated
with $\mathcal{K}^2_+$. Finally, as one of the examples, we make a brief
connection to non-Kramers doublets involved more than a half century 
in works related to Electron paramagnetic resonance or M{\" o}ssbauer
spectroscopy~\cite{EPR_Bleaney-3.14, Stapleton1961, Mueller1968, Telser1998,
Krzystek2006, Maganas2012}, magnetism~\cite{Knizek2014, Hoch2016, Valiev2016},
or conductivity theory~\cite{Boudalis2007, Schaffer2013, Kusunose2016}.

The article is organized as follows. First, we give a general summary on
time-reversal symmetry in the relativistic framework. We then define the
time-reversal generator and formulate the eigenproblem theorem of the
$\mathcal{K}^2_+$ operator, followed by Sec. \ref{sec:K2p} where we prove
this theorem.  In Sec. \ref{sec:paired}, we show a paired structure of the
eigenspectrum.  Subsequently, we discuss a new quantum number of the
Dirac-Coulomb Hamiltonian. Finally, a diagonalization method is employed to
obtain the Kramers configuration state functions of the $\mathcal{K}^2_+$
applied to cases with two (three) open shells. In Appendix \ref{supplement}, we
briefly discuss the cases with four and five unpaired electrons.  In addition,
we provide a simple \texttt{FORTRAN} program able to generate eigenfunctions up to 10
open-shell fermions.


\section{\label{sec:K} Time-reversal symmetry}

Many textbooks on quantum mechanics contain a detailed discussion of
time-reversal symmetry and its applications~\cite{Sachs1987, Lax2001,
DyallFaegri2007}. In the following section, we summarize some of the well
known facts as a starting point for the discussion of the
time-reversal generator in Secs. \ref{sec:Kp}-\ref{sec:K2adapted}.

The one-electron Dirac operator in an external scalar potential $V$ can be written
in atomic units as~\cite{DyallFaegri2007, Schwerdtfeger2002, ReiherWolf2009}
\begin{equation}
  D_i = c \vec\alpha_i\cdot \vec p_i + \beta_i c^2 + V_i,
  \label{eq:dirac}
\end{equation}
where $c$ is the speed of light, $\vec\alpha$ is the off diagonal matrix operator
constructed of gamma matrices in their standard representation
$\vec\alpha=\gamma_0\vec\gamma$, $\vec p = -i \vec\nabla$ is the momentum
operator, and $\beta=\gamma_0$. The subscript $i$ represents the action of the
operators on the $i$th electron.

The Dirac Hamiltonian \eqref{eq:dirac} commutes with an one-electron
time-reversal operator $K_i$, which reflects the fact that the time-dependent Dirac
equation is invariant under time inversion
\begin{equation}
  \left[ D_i, K_i \right ] = 0.
\end{equation}
Fixing the arbitrariness in the phase of the time-reversal operator to $-i$,
$K_i$ can be written as~\cite{Wigner1932, DyallNATO, DyallFaegri2007}
\begin{equation}
  K_i = -i \Sigma_{y,i} K_{0,i},
  \label{eq:TS1}
\end{equation}
where $K_{0,i}$ is the complex conjugation operator and $\Sigma_{y,i}$ is the
four-component spin {\it y} operator expressed via the Pauli matrix
$\sigma_y$ as
\begin{equation}
  \Sigma_y =
    \left(
    \begin{matrix}
        \sigma_y  &  0  \\
        0         &  \sigma_y
    \end{matrix}
    \right).
\end{equation}

The time-reversal operator $K_i$ is an antilinear unitary operator (also called
anti-unitary) satisfying
\begin{gather}
  K_i \left( c_1 \psi + c_2 \phi \right) = c_1^\ast \,K_i\psi + c_2^\ast \,K_i\phi,
  \\
  K_i^\dagger K_i = K_i K_i^\dagger = 1,
  \label{eq:unitarity}
\end{gather}
where $c_1,c_2 \in \mathbb C$ and $\psi, \phi \in [L^2(\mathbb R^3)]^4$ are four-spinors
in the Hilbert space with the inner product
\begin{equation}
  \big< \psi \big| \phi \big> = \int \psi^\dagger \phi \, \mathrm{dV}.
  \label{eq:1e_inner}
\end{equation}
From the definition of an adjoint of antilinear operators
\begin{equation}
  \big< K_i^\dagger \psi \big|     \phi \big> =
  \big<             \psi \big| K_i \phi \big>^\ast
  \label{eq:adjoint}
\end{equation}
it can be shown that the adjoint of the time-reversal operator has the form
\begin{equation}
  K_i^\dagger = i \Sigma_{y,i} K_{0,i},
\end{equation}
which is consistent with the unitary condition in Eq. \eqref{eq:unitarity}.

The closed form for the electron-electron interaction in the relativistic
domain is not known, therefore, only approximate expressions are used. The
commonly applied extensions of one-electron Dirac Hamiltonian to the
many-electron case are the Dirac-Coulomb $H^\mathrm{DC}$ and
Dirac-Coulomb-Breit $H^\mathrm{DB}$ Hamiltonians~\cite{Dyall-chapter5.4}
\begin{gather}
  H^\mathrm{DC} = \sum_i^N D_i + \sum_{i<j}^N \frac{1}{r_{ij}},
  \label{eq:DC}
  \\
  H^\mathrm{DB} = \sum_i^N D_i + \sum_{i<j}^N
    \left[
      \frac{1}{r_{ij}} - \frac{\vec\alpha_i\cdot\vec\alpha_j}{2r_{ij}}
      + \frac{( \vec\alpha_i \cdot \vec r_{ij} ) ( \vec\alpha_j \cdot \vec r_{ij} )}
             {2r^3_{ij}}
    \right].
  \label{eq:DCB}
\end{gather}
Here $N$ is the number of electrons, $\vec r_i$ is the position vector of
the $i$th electron and $r_{ij} = |\vec r_i-\vec r_j|$.

The well-known extension of the one-electron time-reversal operator
\eqref{eq:TS1} to the many-electron case~\cite{Sachs1987,Lax2001} can be
written as
\begin{equation}
  \mathcal{K} = \prod_i^N K_i.
  \label{eq:K2x_def}
\end{equation}
It can be shown that $\mathcal{K}$ is unitary in the sense of
Eq.~\eqref{eq:ne_inner} and commutes with the Hamiltonians \eqref{eq:DC} and
\eqref{eq:DCB}:
\begin{gather}
  \mathcal{K}^\dagger \mathcal{K} =
  \mathcal{K} \mathcal{K}^\dagger = \hat 1,
  \\
  \left[ H^\mathrm{DC}, \mathcal{K} \right ] = 0,
  \label{eq:commut1}
  \\
  \left[ H^\mathrm{DB}, \mathcal{K} \right ] = 0.
  \label{eq:commut2}
\end{gather}
These expressions reflect the fact that $\mathcal{K}$ corresponds to a symmetry
of the system described by the relativistic many-electron Hamiltonians, namely,
time-reversal symmetry. However, $\mathcal{K}$ is antilinear operator, and
thus cannot in general be represented as exponential of a linear Hermitian
operator $\mathcal{O}$, {\it i.e.}, $\mathcal{K} \neq e^{i\mathcal{O}}$.
Therefore, time-reversal symmetry cannot be directly associated with an
observable quantity~\cite{DyallFaegri2007}. In Sec. \ref{sec:K2p}, we show
how the operator $\mathcal{K}$ can be connected to the exponential of an
antilinear operator.

The square of any antilinear operator becomes a linear operator, and in the case
of $\mathcal{K}$, the commutation relations \eqref{eq:commut1} and
\eqref{eq:commut2} of the original operator $\mathcal{K}$ are preserved.
Moreover, due to the simple relation between the one-electron time-reversal
operator and its adjoint
\begin{equation}
  K_i = - K^\dagger_i,
  \label{eq:KvsK+}
\end{equation}
$\mathcal{K}^2$ becomes a Hermitian operator. Thus we can write
\begin{gather}
  \mathcal{K}^2 \left( c_1 \Psi + c_2 \Phi \right)
    = c_1 \,\mathcal{K}^2\Psi + c_2 \,\mathcal{K}^2\Phi,
  \label{eq:K2xlinear}
  \\
  \left[ H^\mathrm{X}, \mathcal{K} \right ] = 0
  \quad \Rightarrow \quad
  \left[ H^\mathrm{X}, \mathcal{K}^2 \right ] = 0,
  \\
  \mathcal{K}^2 = \left( \mathcal{K}^2 \right)^\dagger,
  \label{eq:K2xHermitain}
\end{gather}
where the $N$-electron wave functions $\Psi$ and $\Phi$ belong to the Fock
subspace for $N$ fermions, $\Psi,\Phi \in S^- H^{\otimes N}$, $H=[L^2(\mathbb
R^3)]^4$, with inner product defined in Appendix \ref{app:0}, $\mathrm{X} =
\mathrm{DC},\mathrm{DB}$ and the operator $S^-$ antisymmetrizes a tensor. In
the following, capital Greek letters represent wave functions from the Fock
subspace $S^- H^{\otimes N}$. Finally, utilizing the simple relation for the
one-electron time-reversal operator
\begin{equation}
  K_i^2 = -1,
  \label{eq:Ki2}
\end{equation}
it is straightforward to show that $\mathcal{K}^2$ has the form
\begin{equation}
  \mathcal{K}^2 = (-1)^N \,\hat 1
  \label{eq:K2xform}
\end{equation}
with $\hat 1$ being the identity operator in $S^- H^{\otimes N}$.

Expressions \eqref{eq:K2xlinear}--\eqref{eq:K2xHermitain} define conservation
law for many-electron relativistic systems, with operator of symmetry
\begin{equation}
  e^{i\theta\mathcal{K}^2} = e^{i\theta (-1)^N} \hat1.
\end{equation}
Note that this operator just changes the phase of wave functions. The
corresponding constant of motion
\begin{equation}
  \frac{\mathrm{d}}{\mathrm{dt}}
     \big< \Psi \big| \mathcal{K}^2 \big| \Psi \big> = 0
\end{equation}
represents the fact that the wave functions do not change their boson $(+1)$ or
fermion $(-1)$ symmetry while evolving in time. An equivalent statement is that
Hamiltonians $H^\mathrm{DC}$ and $H^\mathrm{DB}$ share eigenfunctions with the
operator $\mathcal{K}^2$, giving rise to the quantum number $\pm 1$
\cite{Sachs1987,Lax2001}:
\begin{gather}
  H^\mathrm{X} \Psi = E \Psi,
  \label{eq:D_eigen}
  \\
  \mathcal{K}^2 \Psi = (-1)^N \Psi,
  \label{eq:K2x_eigen}
\end{gather}
where $\mathrm{X} = \mathrm{DC},\mathrm{DB}$. Although these are fundamentally
important observations, they are not as useful in spectroscopy as spin
symmetries in the many-electron nonrelativistic domain.

In the next section, we define and describe some properties of the recently
proposed operator $\mathcal{K}^2_+$~\cite{Bucinsky2015, Bucinsky2016}, and
consider in details its relation to $\mathcal{K}$. Unlike the $\mathcal{K}^2$
operator, it has the potential to supplement the role of spin symmetry in the
relativistic domain.


\section{\label{sec:Kp} Time-reversal generator}

Bu{\v c}insk{\' y} {\it et al.}~\cite{Bucinsky2015, Bucinsky2016} recently
proposed the many-electron operator
\begin{equation}
  \mathcal{K}_+=\sum_i^N K_i.
  \label{eq:K2p_def}
\end{equation}
The operator in Eq. \eqref{eq:K2p_def} is antilinear, but unlike $\mathcal{K}$ it is not
unitary [for the definition of $\mathcal{K}_+^\dagger$, see Eq. \eqref{eq:ne_inner}]:
\begin{equation}
  \mathcal{K}_+^\dagger \mathcal{K}_+ =
  \mathcal{K}_+ \mathcal{K}_+^\dagger \neq 1,
  \label{eq:K+dag_K+_neq_1}
\end{equation}
thus, it does not represent a symmetry operation. However, as it is shown
in Sec. \ref{sec:K2p}, it is connected to $\mathcal{K}$ through
\begin{equation}
  \mathcal{K} = e^{ \frac{\pi}{2} \mathcal{K}_+ }.
  \label{eq:relation}
\end{equation}
This exponential mapping is in some aspects similar to the standard relation
between a unitary operator $\mathcal{U}$ and the corresponding Hermitian
operator $\mathcal{O}$, $\mathcal{U}=e^{i\mathcal{O}}$. If such operators
commute with the Hamiltonian, the symmetry and conservation law of the system
are defined by $\mathcal{U}$ and $\mathcal{O}$, respectively. Thus, we call the
operator $\mathcal{K}_+$ a generator of the time-reversal operator
$\mathcal{K}$, in analogy to the elements of the Lie algebra being
infinitesimal generators of the Lie group.  Unfortunately, $\mathcal{K}_+$ is
still an antilinear operator, and it can therefore not be associated with an
observable quantity (it is not diagonalizable) even though it commutes with the
Dirac-Coulomb Hamiltonian
\begin{gather}
  \left[ H^\mathrm{DC}, \mathcal{K}_+ \right ] = 0,
  \\
  \left[ H^\mathrm{DB}, \mathcal{K}_+ \right ] \neq 0.
\end{gather}
For completeness, we also note that $\mathcal{K}_+$ does not commutate with the
Dirac-Coulomb-Breit Hamiltonian. However, it can be shown that its square is
a linear Hermitian operator [utilizing Eq.~\eqref{eq:KvsK+}] which still
commutes with the Dirac-Coulomb Hamiltonian:
\begin{gather}
  \mathcal{K}^2_+ \left( c_1 \Psi + c_2 \Phi \right)
    = c_1 \,\mathcal{K}^2_+\Psi + c_2 \,\mathcal{K}^2_+\Phi,
  \label{eq:K2plinear}
  \\
  \left[ H^\mathrm{DC}, \mathcal{K}_+ \right ] = 0
  \quad \Rightarrow \quad
  \left[ H^\mathrm{DC}, \mathcal{K}^2_+ \right ] = 0,
  \label{eq:K2pcommute}
  \\
  \mathcal{K}^2_+ = \left( \mathcal{K}^2_+ \right)^\dagger.
  \label{eq:K2pHermitain}
\end{gather}
Therefore, it corresponds to an observable and as it turns out it has a much
richer eigenvalue spectrum than $\mathcal{K}^2$ [Eq. \eqref{eq:K2x_eigen}]. In
Sec. \ref{sec:K2p}, we prove the following eigenvalue theorem:
\begin{equation}
  \begin{gathered}
  \mathcal{K}_+^2 \Psi = -k^2 \Psi, \qquad k\in\mathbb{N}_0,
  \\
  \begin{array}{ccc}
      \text{odd }  N & \Leftrightarrow & \text{odd } k,
      \\
      \text{even } N & \Leftrightarrow & \text{even } k,
  \end{array}
  \end{gathered}
  \label{eq:K2p_eigen1}
\end{equation}
Here, $N$ refers to the total number of electrons in a system. When
constructing the eigenfunction $\Psi$ of $\mathcal{K}_+^2$ as a specific linear
combination of Slater determinants, where each determinant is composed of
Kramers-restricted molecular orbitals (see Sec. \ref{sec:K2adapted}), we
observe that these eigenfunctions have a more refined eigenvalue spectrum and
degeneracy
\begin{equation}
  \begin{gathered}
  \mathcal{K}_+^2 \Psi(N,N_O) = -k^2 \Psi(N,N_O),
  \\
  \begin{array}{ccc}
      \text{odd }  N & \Leftrightarrow & k = 1,3, \cdots , N_O,
      \\
      \text{even } N & \Leftrightarrow & k = 0,2, \cdots , N_O,
  \end{array}
  \end{gathered}
  \label{eq:K2p_eigen2}
\end{equation}
where $N_O$ is the number of unpaired electrons (open shells).

Note that in Eqs.~\eqref{eq:K2p_eigen1} and \eqref{eq:K2p_eigen2}, as well as
in the following discussion, we omit the index $k$ for the eigenfunction
$\Psi_k$ to simplify the notation. However, the reader should keep in mind that
eigenfunctions $\Psi$ always associate with a specific eigenvalue $-k^2$.


\section{\label{sec:K2p} Eigenspectrum of $\mathcal{K}^2_+$ operator}

In this section, we prove the eigenvalue theorem \eqref{eq:K2p_eigen1} and
establish the relation between the many-electron operators $\mathcal{K}$
[Eq.~\eqref{eq:K2x_def}] and $\mathcal{K}_+$ [Eq.~\eqref{eq:K2p_def}].

Since the operators $\mathcal{K}^2_+$ and $\mathcal{K}^2$ commute
\begin{equation}
  \left[ \mathcal{K}^2_+, \mathcal{K}^2 \right] = 0
\end{equation}
and $\mathcal{K}^2$ is just a scaled identity operator [Eq.
\eqref{eq:K2xform}], both operators share the same set of eigenfunctions
\begin{gather}
  \mathcal{K}_+^2 \Psi = \kappa \Psi,
  \label{eq:K2p_gen}
  \\
  \mathcal{K}^2 \Psi = (-1)^N \Psi.
  \label{eq:K2x_gen}
\end{gather}
Here, the eigenvalue $\kappa$ is considered an unknown real number. In the
following, we show that the form of eigenvalues $\kappa$ [Eq.~\eqref{eq:K2p_eigen1}]
is a direct consequence of expressions \eqref{eq:K2p_gen} and
\eqref{eq:K2x_gen}.

Defining $e^{\theta K_i}$ through the Taylor series expansion and utilizing
repeatedly the property of the one-electron time-reversal operator
\eqref{eq:Ki2}, we can write (see Appendix \ref{app:1})
\begin{equation}
  e^{\theta K_i} = \cos (\theta) + K_i \, \sin(\theta).
  \label{eKi}
\end{equation}
Choosing $\theta=\pi/2$ or $\theta=\pi$, we get
\begin{gather}
  e^{\frac{\pi}{2} K_i}=K_i,
  \label{eq:Kpi2}
  \\
  e^{\pi K_i}=-1.
  \label{eq:Kpi}
\end{gather}
Employing Eq.~\eqref{eq:Kpi2} and the commutation relation $[K_i,K_j]=0$ in a
many-fermion case, we can rewrite the time-reversal operator~\eqref{eq:K2x_def}
as
\begin{equation}
    \mathcal{K} = \prod_i^N K_i = \prod_i^N e^{\frac{\pi}{2} K_i}
    = e^{\frac{\pi}{2} \sum_i^N K_i}.
  \label{eq:KxandKp}
\end{equation}
As a result, we obtain the relation \eqref{eq:relation} for many-fermion operators
\begin{equation}
  \mathcal{K} = e^{\frac{\pi}{2} \mathcal{K}_+}
  \label{eq:Kx_vs_Kp}
\end{equation}
and similarly for their adjoint
\begin{equation}
  \mathcal{K}^\dagger = e^{\frac{\pi}{2} \mathcal{K}_+^\dagger}.
\end{equation}
Multiplying the last two equations and realizing that $\mathcal{K}_+^\dagger =
-\mathcal{K}_+$ [see Eq. \eqref{eq:KvsK+}], the unitarity of $\mathcal{K}$
can readily be obtained.

We now turn our attention to the square of the time-reversal operator $\mathcal{K}$:
\begin{equation}
  \mathcal{K}^2 = e^{\pi \mathcal{K}_+}
\end{equation}
since it provides us with the link between the eigenvalues of
$\mathcal{K}^2$ and $\mathcal{K}_+^2$ operators. Applying
Eq.~\eqref{eq:K2p_gen}, we obtain
\begin{equation}
  \begin{aligned}
  e^{\pi \mathcal{K}_+}\Psi
  & = \left(1 + \pi \mathcal{K}_+ + \frac{\pi^2}{2!} \mathcal{K}_+^2 + \cdots\right) \Psi \\
  & = \left(1   + \frac{\pi^2}{2!}\kappa + \frac{\pi^4}{4!}\kappa^2 + \cdots\right)\Psi \\
  & + \left(\pi + \frac{\pi^3}{3!}\kappa + \frac{\pi^5}{5!}\kappa^2 + \cdots\right) \mathcal{K}_+ \Psi,
  \end{aligned}
\end{equation}
and Eq.~\eqref{eq:K2x_gen}, we get
\begin{equation}
  e^{\pi \mathcal{K}_+}\Psi = \mathcal{K}^2 \Psi = (-1)^{N} \Psi.
\end{equation}
Combining the last two equations, we can write
\begin{equation}
  \begin{aligned}
  (-1)^{N} \Psi
  & = \left(  1 + \frac{\pi^2}{2!}\kappa + \frac{\pi^4}{4!}\kappa^2 + \cdots\right) \Psi \\
  & + \left(\pi + \frac{\pi^3}{3!}\kappa + \frac{\pi^5}{5!}\kappa^2 + \cdots\right)
  \mathcal{K}_+ \Psi.
  \end{aligned}
\label{eq:kappa}
\end{equation}

It is worth to examine two possibilities of the action of
$\mathcal{K}_+$ on the wave function $\Psi$:
\begin{gather}
  \mathcal{K}_+ \Psi = 0,
  \label{eq:KpPsi1}
  \\
  \mathcal{K}_+ \Psi \equiv \Phi.
  \label{eq:KpPsi2}
\end{gather}
In the first case, by substituting Eq.~\eqref{eq:KpPsi1} into
Eqs.~\eqref{eq:K2p_gen} and \eqref{eq:kappa}, we immediately see that
Eq.~\eqref{eq:KpPsi1} is satisfied only for boson-type wave functions
(even number of electrons)
\begin{equation}
  \mathcal{K}_+ \Psi = 0 \quad \Rightarrow \quad
  \kappa=0               \quad \Rightarrow \quad
  (-1)^{N} \Psi = \Psi.
  \label{eq:Kp_0}
\end{equation}
In the second case, it is possible to show that the wave function $\Phi$ is
orthogonal to $\Psi$, but since $\mathcal{K}_+$ is not unitary, $\Phi$
is not normalized to one (for proof, see Appendix \ref{app:2}):
\begin{align}
  \big< \Psi \big| \Phi \big> &= 0,
  \label{eq:ortho}
  \\
  \big< \Phi \big| \Phi \big> &= -\kappa.
  \label{eq:norm}
\end{align}

Integrating Eq. \eqref{eq:kappa} with $\left<\Psi\right|$ and $\left<\Phi\right|$
we get two expressions
\begin{align}
  (-1)^{N} &=  1 + \frac{\pi^2}{2!}\kappa + \frac{\pi^4}{4!}\kappa^2 + \cdots,
  \label{eq:expr1}
  \\
  0 &= \pi\kappa + \frac{\pi^3}{3!}\kappa^2 + \frac{\pi^5}{5!}\kappa^3 + \cdots.
  \label{eq:expr2}
\end{align}
Since $\kappa$ is an eigenvalue of a Hermitian operator it must be real,
and we can therefore examine $\kappa$ being a positive or a negative real number.
For this purpose, we use the ansatz $\kappa=k^2$ and $\kappa=-k^2$,
respectively.

In the case of $\kappa=k^2$, we get from Eqs. \eqref{eq:expr1} and
\eqref{eq:expr2}
\begin{align}
  (-1)^{N} &= \cosh(\pi k),
  \\
  0 &= k \sinh(\pi k),
\end{align}
which is satisfied only for $k=0$ and even number of electrons $N$. More
interesting is the case of $\kappa=-k^2$, where we get
\begin{align}
  (-1)^{N} &= \cos(\pi k),
  \\
  0 &= k \sin(\pi k),
\end{align}
which is satisfied for integer numbers ($k \in \mathbb{Z}$) with the following
rule:
\begin{align}
  \begin{array}{ccc}
    \text{odd }  N & \Leftrightarrow & \text{odd } k,
    \\
    \text{even } N & \Leftrightarrow & \text{even } k,
  \end{array}
\end{align}
We can further restrict $k$ to positive integers including zero $k \in
\mathbb{N}_0$ since both positive and negative $k$ produce the same eigenvalues
$\kappa$. Thus, we have proved the theorem \eqref{eq:K2p_eigen1}.


\section{\label{sec:paired} Paired eigenfunctions of the $\mathcal{K}^2_+$ operator}

In Sec. \ref{sec:K2p}, we have seen that the time-reversal generator when
acting on normalized wave functions $\Psi$ produces a non-normalized wave
function $\Phi$ [see Eqs. \eqref{eq:KpPsi2} and \eqref{eq:norm}]. By choosing
the definition in Eq. \eqref{eq:KpPsi2} to
\begin{equation}
  \mathcal{K}_+ \Psi \equiv k \widetilde\Psi
  \quad \Rightarrow \quad
  \big< \widetilde\Psi \big| \widetilde\Psi \big> = 1
  \label{eq:pseudo_pair1}
\end{equation}
such that $k>0$ and $-k^2$ is the eigenvalue of $\Psi$ defined in
Eq.~\eqref{eq:K2p_eigen1}, then $\widetilde\Psi$ is a normalized wave function.
Applying operator $\mathcal{K}_+$ on Eq.~\eqref{eq:pseudo_pair1} and employing
Eq.~\eqref{eq:K2p_eigen1} we get
\begin{equation}
  \mathcal{K}_+ \widetilde\Psi = - k \Psi.
  \label{eq:pseudo_pair2}
\end{equation}
Note that the choice of right-hand side in Eq.~\eqref{eq:pseudo_pair1} fixes the
relative phase of wave functions $\Psi$ and $\widetilde\Psi$.
The relations \eqref{eq:pseudo_pair1} and \eqref{eq:pseudo_pair2} between
$\Psi$ and $\widetilde\Psi$ have been observed previously for Kramers-restricted
Slater determinants~\cite{Bucinsky2016}.

The wave functions $\Psi$ and $\widetilde\Psi$ are normalized, orthogonal, and
share the same eigenvalue (see Appendix \ref{app:2})
\begin{equation}
  \begin{aligned}
  \mathcal{K}_+^2 \Psi = -k^2 \Psi,
  \\
  \mathcal{K}_+^2 \widetilde\Psi = -k^2 \widetilde\Psi.
  \end{aligned}
  \label{eq:pair_eigen}
\end{equation}
The only exception arises for $k=0$, for which $\mathcal{K}_+\Psi$
is zero and thus $\widetilde\Psi$ is not uniquely defined.
Nevertheless, due to Eqs. \eqref{eq:KpPsi2} and \eqref{eq:norm}, we can change
the implication in expression \eqref{eq:Kp_0} to an equivalence
\begin{equation}
  \left[ \, \mathcal{K}_+ \Psi = 0
            \quad \Leftrightarrow \quad
            k=0                          \, \right]
  \quad \Rightarrow \quad
  (-1)^{N} \Psi = \Psi.
  \label{eq:k_zero}
\end{equation}

As a result, for $k\neq0$ the eigenspectrum of $\mathcal{K}_+^2$ is at least
two times degenerate, where Eqs. \eqref{eq:pseudo_pair1} and
\eqref{eq:pseudo_pair2} describe the connection between these degenerate wave
functions.  The pair structure \eqref{eq:pseudo_pair1}--\eqref{eq:pair_eigen} is
similar to the Kramers pairs arising from the time-reversal symmetry operator
$\mathcal{K}$, defined as
\begin{equation}
  \mathcal{K} \Psi \equiv \overbar\Psi.
\end{equation}
From the form of the $\mathcal{K}^2$ operator \eqref{eq:K2xform}, it is clear
that $\overbar\Psi$ has the same eigenvalue as $\Psi$ and because $\mathcal{K}$
is unitary, $\overbar\Psi$ remains normalized. Indeed, there is a close
connection between these two paired structures (see Appendix \ref{app:3}),
where for $k=0$
\begin{equation}
  \mathcal{K}_+\Psi = 0
  \quad \wedge \quad
  \Psi = \overbar\Psi
\end{equation}
and for $k\neq0$
\begin{equation}
  \def\arraystretch{1.3}
  \left(
  \begin{array}{rr}
       \cos\big(\frac{\pi}{2}k\big)  &  \sin\big(\frac{\pi}{2}k\big)  \\
      -\sin\big(\frac{\pi}{2}k\big)  &  \cos\big(\frac{\pi}{2}k\big)
  \end{array}
  \right)
  \left(
  \begin{array}{c}
       \Psi \\
       \widetilde\Psi
  \end{array}
  \right)
  =
  \left(
  \begin{array}{c}
       \overbar\Psi \\
       \overbar{\widetilde\Psi}
  \end{array}
  \right).
  \label{eq:paired_structures1}
\end{equation}
In addition, because $k$ is an integer, we arrive at the following two cases:
\begin{equation}
  \begin{aligned}
    \text{even }k \quad \Rightarrow \quad
    &
    \cos\left(\frac{\pi}{2}k\right)
    \left(
    \begin{array}{c}
         \Psi \\
         \widetilde\Psi
    \end{array}
    \right)
    =
    \left(
    \begin{array}{c}
         \overbar\Psi \\
         \overbar{\widetilde\Psi}
    \end{array}
    \right),
  \\
    \text{odd }k \quad \Rightarrow \quad
    &
    \sin\left(\frac{\pi}{2}k\right)
    \left(
    \begin{array}{c}
         \widetilde\Psi\\
         -\Psi
    \end{array}
    \right)
    =
    \left(
    \begin{array}{c}
         \overbar\Psi \\
         \overbar{\widetilde\Psi}
    \end{array}
    \right).
  \end{aligned}
  \label{eq:paired_structures2}
\end{equation}
The sine and cosine functions change only the sign of the wave functions, and
can be ignored in the following discussion. Based on Eqs.
\eqref{eq:paired_structures2}, we conclude that for boson systems (even $k$ and
$N$), the time-reversal operator produces the same wave function and for a
fermion system (odd $k$ and $N$), $\overbar\Psi$ and $\widetilde\Psi$ are equal.
In other words, the wave function $\widetilde\Psi$ in the eigenspectrum of
$\mathcal{K}_+^2$ can be reached by operating both with $\mathcal{K}_+$ and
$\mathcal{K}$ in the fermion case, but only with the $\mathcal{K}_+$ operator
in the boson case.


\section{\label{sec:DC} Quantum number of the many-electron Dirac-Coulomb Hamiltonian}

In the case of \emph{infinite-dimensional} Hilbert spaces, two operators which
commute do not in general produce the same set of eigenfunctions, and therefore
wave functions $\Psi$ in Eq.~\eqref{eq:K2p_eigen1} are not necessarily
eigenfunctions of the Dirac-Coulomb Hamiltonian, despite of the commutation
relation \eqref{eq:K2pcommute}. The existence of a common set of degenerate
eigenfunctions of two commuting operators must be proved for each case
separately. For example, the $\mathcal{K}^2$ operator has the very simple form
\eqref{eq:K2xform}, and thus it is easy to see that it shares eigenfunctions with
the many-electron relativistic Hamiltonians [Eqs.~\eqref{eq:D_eigen} and
\eqref{eq:K2x_eigen}] and $\mathcal{K}^2_+$ operator
[Eqs.~\eqref{eq:K2p_gen} and \eqref{eq:K2x_gen}].

The form of the $\mathcal{K}^2_+$ is not as trivial as $\mathcal{K}^2$, and the
proof that the former operator shares eigenfunctions with the Dirac-Coulomb
Hamiltonian is not known to the authors, despite the fact that they commute
\eqref{eq:K2pcommute}.  However, if two operators commute in
\emph{finite-dimensional} Hilbert space it can be shown that they automatically
share the same set of eigenvectors. Fortunately, this can be utilized for
$\mathcal{K}^2_+$ and $H^\mathrm{DC}$ operators when represented in Fock
subspace $F(M,N)$ (subspace of $S^- H^{\otimes N}$):
\begin{equation}
  \left[ \bm{H}^\mathrm{DC}, \bm{\mathcal{K}}^2_+ \right ] = 0,
  \label{eq:DCvsK2_FNM}
\end{equation}
where a finite-dimensional basis in $F(M,N)$, used to represent the operators
in Eq.~\eqref{eq:DCvsK2_FNM}, contains all Kramers-restricted Slater
determinants (KRSD) obtained by distributing $N$ electrons among $M$ four-spinors.
The proof of relation~\eqref{eq:DCvsK2_FNM}, which assumes the use of
an orthonormal restricted kinetically balanced basis~\cite{Stanton} to
represent four-spinors, is given in Appendix \ref{app:4}.  Finally, we can
conclude that commutation relation \eqref{eq:DCvsK2_FNM} leads to a new
quantum number \eqref{eq:K2p_eigen1} for solutions of the Dirac-Coulomb
Hamiltonian.

In the nonrelativistic theory, multi-configuration post-Hartree--Fock methods
often utilize linear combination of Slater determinants (configuration state
functions~\cite{Helgaker-2.5}) as many-electron basis. These functions
account for spin symmetry of the one-component Hamiltonians (being
eigenfunctions of spin operators), potentially reducing the computational cost
and simplifying the analysis of the solutions. In the relativistic domain, the spin
symmetry is broken but since $\bm{\mathcal{K}}^2_+$ and $\bm{H}^\mathrm{DC}$
share the same set of eigenvectors, the relativistic counterpart of
configuration state functions can be defined using eigenvectors of the
$\bm{\mathcal{K}}^2_+$ operator instead. We denote these functions as
Kramers configuration state functions (KCSF) and discuss their construction in
Sec.~\ref{sec:K2adapted}. The biggest potential advantage of these functions
as a many-electron basis lies in the ability to predict the structure of
the matrix representation of the Dirac-Coulomb Hamiltonian and other operators
$\mathcal{O}$ in this basis. This can give rise to selection rules (forbidden
or allowed transition) in the relativistic domain. However, a detailed study of
such rules is beyond the scope of this work. Herein, we show only four
simple examples.

As a first example, let us consider Hermitian time-reversal anti-symmetric
operators
\begin{equation}
  \begin{gathered}
    \mathcal{O}^\dagger = \mathcal{O},
    \\
    \mathcal{K}^\dagger \mathcal{O} \mathcal{K} = - \mathcal{O}.
  \end{gathered}
  \label{eq:mag_field}
\end{equation}
A typical example are operators responsible for interaction with magnetic
fields or operators for total spin. For an even number of electrons, it holds
that the Kramers-partner wave function $\overbar\Psi$ is equal, up to a sign, to
the wave function $\Psi$ [see Eq.~\eqref{eq:paired_structures2}]. It is then
easy to show that the inner product of this wave function with the operator
$\mathcal{O}$ is zero (see Appendix \ref{app:5}):
\begin{equation}
  \text{even } N \quad \Rightarrow \quad
  \big< \Psi \big| \mathcal{O} \big| \Psi \big> = 0.
  \label{eq:no_spin}
\end{equation}
Although this expression holds for an exact wave function of the Dirac-Coulomb
Hamiltonian, it can be extended also to KCSF.  As a consequence, the diagonal
elements of magnetic field operators represented in the KCSF basis are strictly
zero for even-electron systems.

Another example is the connection of paired wave functions $\Psi$ and
$\widetilde\Psi$ [Eq.~\eqref{eq:pseudo_pair1}] with eigenvalues of the Dirac-Coulomb
Hamiltonian. In the case of $k\ne0$, when combining the commutation relation
\eqref{eq:DCvsK2_FNM} with the definition of the paired wave function
\eqref{eq:pseudo_pair1} we obtain
\begin{align}
  \bm{H}^\mathrm{DC} \bm{C} = E \bm{C},
  \\
  \bm{H}^\mathrm{DC} \bm{\widetilde C} = E \bm{\widetilde C},
\end{align}
where $\bm{C}$ are expansion coefficients in the KRSD basis. The paired
eigenvectors thus share the same eigenvalue of both Dirac-Coulomb and
$\bm{\mathcal{K}}^2_+$ operators [see also Eq. \eqref{eq:pair_eigen}]. In
other words, if an eigenvector of the $\bm{H}^\mathrm{DC}$ Hamiltonian is
associated with a non-zero quantum number $-k^2$, its energy level is at least
two times degenerate. Consequently, if an eigenvector $\bm{C}^{nd}$ is
non-degenerate then its quantum number $-k^2$ is equal to zero, which may
happen only for systems with an even number of electrons [Eq.~\eqref{eq:k_zero}],
{\it i.e.},
\begin{equation}
  \begin{gathered}
    \bm{H}^\mathrm{DC} \bm{C}^{nd} = E^{nd} \bm{C}^{nd}
    \\
    \Rightarrow \quad
    \bm{\mathcal{K}}_+ \bm{C}^{nd} = 0
    \quad \Leftrightarrow \quad
    \text{even } N.
  \end{gathered}
\end{equation}
Similarly, all energy levels are at least $2n$ times degenerate
($n\in\mathbb{N}$) for systems with an odd number of electrons, which is
a well-known fact easy to prove facilitating the time-reversal operator
$\mathcal{K}$~\cite{Rosch1980}. We can also translate these statements to the
nonrelativistic framework, where the energetically non-degenerate states are
allowed only for systems with an even number of electrons, like for instance
closed-shell or open-shell singlet states.

As a third example, let us consider a doubly degenerate states $\{\Psi_1,
\Psi_2\}$ of the Dirac-Coulomb Hamiltonian. According to the above discussion,
these states have the same eigenvalue $-k^2$ and behave under time-reversal
symmetry as [see Eq. \eqref{eq:paired_structures2}]
\begin{align}
    \text{even }N \quad \Rightarrow \quad
    &
    \mathcal{K}\Psi_i = \pm \Psi_i,
    \quad i = 1,2,
    \label{eq:non_kramer_2}
  \\
    \text{odd }N \quad \Rightarrow \quad
    &
    \mathcal{K}\Psi_1 = \pm \Psi_2.
    \label{eq:kramer_2}
\end{align}
The matrix representation of any Hermitian operator in the basis of two wave
functions, either degenerate as $\{\Psi_1,\Psi_2\}$ or non-degenerate, can
be expanded as a linear combination of the identity matrix and the Pauli
matrices. For Hermitian time-reversal antisymmetric
operators~\eqref{eq:mag_field}, only Pauli matrices contribute in the case of
so-called Kramers doublet~\eqref{eq:kramer_2}, and only the Pauli {\it y}
matrix contributes in the case of so-called non-Kramers
doublet~\eqref{eq:non_kramer_2}~\cite{EPR_Bleaney-3.14}. To prove the last
statement, use expressions \eqref{eq:non_kramer_2} and \eqref{eq:kramer_2} and
techniques from Appendix \ref{app:5}.

As a fourth example, we consider an even-electron system and two arbitrary
states of the Dirac-Coulomb Hamiltonian, $\Psi_1$ and $\Psi_2$. These two wave
functions are related by time-reversal symmetry via
Eq.~\eqref{eq:non_kramer_2}, thus, the expectation value of an operator
$\mathcal{O}$ [Eq. \eqref{eq:mag_field}] can be expressed as
\begin{equation}
      \big<            \Psi_1 \big| \mathcal{O} \big|            \Psi_2 \big> =
  \pm \big< \mathcal{K}\Psi_1 \big| \mathcal{O} \big| \mathcal{K}\Psi_2 \big> =
  \mp \big<            \Psi_1 \big| \mathcal{O} \big|            \Psi_2 \big>^\ast.
  \label{eq:mat_element}
\end{equation}
Therefore, if wave functions $\Psi_1$ and $\Psi_2$ transform under time-reversal
symmetry with the same (different) sign the matrix element~\eqref{eq:mat_element}
is a pure imaginary (real) number. As a result for systems with even number of
electrons the matrix representations of operators responsible for magnetic
interactions are either pure real or pure imaginary numbers on off diagonal
\eqref{eq:mat_element} and zero on the diagonal~\eqref{eq:no_spin}. This can
help to design effective spin Hamiltonians used to characterize the
heavy-element containing systems.

Finally, we note that the previous findings also hold for any approximate
two-component Hamiltonians involving the Coulomb operator for electron-electron
interaction. All those Hamiltonians commute with two-component version of the
$\bm{\mathcal{K}}^2_+$ operator and share the same set of eigenvectors. 
Also note that a time-reversal antisymmetric magnetic field operator breaks the
commutation relation \eqref{eq:DCvsK2_FNM} since the $\mathcal{K}_+^2$ operator
has no special commutation or anti-commutation relation with the magnetic field
operator.


\section{\label{sec:K2adapted} Kramers configuration state functions}

In the previous section, we have argued that in order to relate eigenfunctions
of the Dirac-Coulomb and $\mathcal{K}_+^2$ operators and thus introducing a
new quantum number, we need to represent both operators in the Fock subspace
$F(M,N)$.  Moreover, in any practical application of the quantum theory, the
discretization of the infinite-dimensional problems is essential. In this
section, we investigate the matrix representation of the $\mathcal{K}_+^2$
operator in the Fock subspace $F(M,N)$ in more details. The complete basis in
$F(M,N)$ contains all Kramers-restricted Slater determinants obtained by
distributing $N$ electrons among $M$ four-spinors.  Since a one-electron
operator $K_i$, alike spin operators in the nonrelativistic theory, mixes only
associated Kramers pairs, when investigating the $\bm{\mathcal{K}}_+^2$
eigenvectors it is sufficient to involve only determinants with a constant
number of excitations ({\it i.e.}, constant number of unpaired electrons). In
other words, $\bm{\mathcal{K}}_+^2$ has a block-diagonal structure in the Fock
subspace $F(M,N)$. Here, the reader is referred to Appendix \ref{app:6}, where
the form of the $\mathcal{K}_+^2$ operator in the second quantization formalism
is employed to prove this statement.

Let us consider a basis consisting of Kramers-restricted Slater determinants
with $N_O$ unpaired (open-shell) electrons $\{\Phi_i(N_O)\}$. Each Slater
determinant is constructed from a set of Kramers paired four-spinors
\cite{Dyall-chapter9}. To indicate the Kramers paired structure of the spinor
$m$, we use bar over the index ($\overbar m$). The $\mathcal{K}_+^2$ operator
for an $N$-electron system can be expressed as
\begin{equation}
  \mathcal{K}_+^2 = -N \hat{1} + 2 \sum_{i<j}^N K_i K_j
  \label{eq:Kp2_explicit}
\end{equation}
and the definition of the matrix elements of $\mathcal{K}_+^2$ in the
$\{\Phi_i(N_O)\}$ basis reads as
\begin{equation}
  \left( \mathcal{K}_+^2 \right)_{ij} =
    \big< \Phi_i \big| \mathcal{K}_+^2 \big| \Phi_j \big>.
  \label{eq:Kp2_mat}
\end{equation}
From the form of the $\mathcal{K}_+^2$ operator~\eqref{eq:Kp2_explicit} and the
discussion in the previous sections, we can draw some general conclusions about
the properties of the matrix elements~\eqref{eq:Kp2_mat}:
\begin{enumerate}[label=(\roman*)]
\item
  The matrix elements of $\mathcal{K}_+^2$ operator in the basis of
  Kramers-restricted wave functions are real numbers, as can be easily seen from
  the second quantization form of $\mathcal{K}_+$ (see Appendix \ref{app:6}).
\item
  The diagonal elements have the simple form (see also Ref.~\cite{Bucinsky2015})
  \begin{equation}
    \left( \mathcal{K}_+^2 \right)_{ii} = -N_O.
  \end{equation}
\item
  The $\{\Phi_i(N_O)\}$ manifold can be split into two sets based on the even
  ({\it e}) and odd ({\it o}) number of unpaired barred spinors in the determinants,
  $\{\Phi^e_i\}$ and $\{\Phi^o_i\}$, respectively. The inner product between these
  two sets is zero because $\mathcal{K}_+^2$ contains either double $K_i K_j$ or
  neutral $K_i K_i=-1$ contributions [see Eq. \eqref{eq:Kp2_explicit}]
  \begin{equation}
     \big< \Phi^e_i \big| \mathcal{K}_+^2 \big| \Phi^o_j \big> = 0.
     \label{eq:block_diag}
  \end{equation}
\item
  For system with an odd number of electrons it holds
  \begin{equation}
     \text{odd }N \quad \Rightarrow \quad
     \big< \Phi^o_i \big| \mathcal{K}_+^2 \big| \Phi^o_j \big> =
     \big< \Phi^e_i \big| \mathcal{K}_+^2 \big| \Phi^e_j \big>.
     \label{eq:block_equal}
  \end{equation}
  To prove expression~\eqref{eq:block_equal}, one needs the connection between
  even and odd sets
  \begin{equation}
    \Phi^o_i = \mathcal{K} \Phi^e_i
    \label{eq:even_odd}
  \end{equation}
  techniques from Appendix~\ref{app:2}, the commutation relation
  $[\mathcal{K}_+,\mathcal{K}]=0$, unitarity of $\mathcal{K}$, and
  real-valued matrix elements $\left( \mathcal{K}_+^2 \right)_{ij}$.
\item
  For $k\ne0$, eigenvectors of $\big< \Phi^x_i \big| \mathcal{K}_+^2 \big|
  \Phi^x_j \big>$ corresponding to the even ($x=e$) or odd ($x=o$) set are
  related to each other by the $\mathcal{K}_+$ operator (see discussion in Sec.
  \ref{sec:paired}).
\end{enumerate}

To build the Kramers configuration state functions one needs to diagonalize the
matrix representation of the $\mathcal{K}_+^2$ operator in the basis of
Kramers-restricted Slater determinants $\{\Phi_i(N_O)\}$, where $N_O$ denotes a
particular number of unpaired electrons. Intermediate products $\mathcal{K}_+^2
\Phi_j$ for two and three unpaired electrons are listed in
Appendix~\ref{app:7}. Here, we only summarize and discuss the final matrix
elements \eqref{eq:Kp2_mat}.

For two unpaired electrons, the basis consists of $\{ \Phi_{12},
\Phi_{\bar{1}\bar{2}}, \Phi_{1\bar{2}}, \Phi_{\bar{1}2} \}$, and the
matrix elements of the $\mathcal{K}_+^2$ operator are
\begin{equation}
\begin{aligned}
\begin{array}{rrrrrrr}
              &                  &{1}{2}&\bar{1}\bar{2}&{1}\bar{2}&\bar{1}{2}& \\
{1}{2}        &\ldelim({4}{0.5em}&-2    &    2         &  0       &   0      &\rdelim){4}{0.5em} \\
\bar{1}\bar{2}&                  & 2    &   -2         &  0       &   0      & \\
    {1}\bar{2}&                  & 0    &    0         &  -2      &  -2      & \\
\bar{1}{2}    &                  & 0    &    0         &  -2      &  -2      & \\
\end{array}
\end{aligned}.
\label{eq:K2+ij2M}
\end{equation}
As discussed earlier, the matrix \eqref{eq:K2+ij2M} is real and has a block-diagonal
structure with the diagonal elements equal to minus the number of unpaired
electrons. After diagonalization of the matrix \eqref{eq:K2+ij2M} we
obtain the eigenvalues $e_i$ and eigenfunctions $\Psi_i$ in the following form:
\begin{equation}
  \begin{aligned}
  \Psi_1 = \tfrac{1}{\sqrt{2}} \left( \Phi_{12} - \Phi_{\bar{1}\bar{2}} \right), &\qquad e_1=-4,
  \\
  \Psi_2 = \tfrac{1}{\sqrt{2}} \left( \Phi_{1\bar{2}} + \Phi_{\bar{1}2} \right), &\qquad e_2=-4,
  \\
  \Psi_3 = \tfrac{1}{\sqrt{2}} \left( \Phi_{12} + \Phi_{\bar{1}\bar{2}} \right), &\qquad e_3= 0,
  \\
  \Psi_4 = \tfrac{1}{\sqrt{2}} \left( \Phi_{1\bar{2}} - \Phi_{\bar{1}2} \right), &\qquad e_4= 0.
  \end{aligned}
\label{eq:K2+ij2eval}
\end{equation}
The wave functions in Eqs.~\eqref{eq:K2+ij2eval} satisfy expressions
\eqref{eq:K2p_eigen2}, \eqref{eq:no_spin}, and \eqref{eq:mat_element}. The
paired structure for system with an even number of electrons, as described in
Sec. \ref{sec:K2p}, can also be readily verified:
\begin{equation}
  \def\arraystretch{1.5}
  \begin{array}{ll}
  \mathcal{K}_+ \Psi_1 = 2 \Psi_2,      & {}
  \\
  \mathcal{K}_+ \Psi_2 = -2 \Psi_1,     & {}
  \\
  \mathcal{K}_+ \Psi_i = 0,             & i = 3,4,
  \\
  \mathcal{K} \Psi_i = - \Psi_i, & i = 1,2,
  \\
  \mathcal{K} \Psi_i =   \Psi_i, & i = 3,4.
  \end{array}
\end{equation}
Thus according to notation in Sec. \ref{sec:K2p} we can write
\begin{equation}
  \Psi_2 = \widetilde\Psi_1.
\end{equation}

For comparison, let us consider the three-electron open-shell case. Because the
matrix~\eqref{eq:Kp2_mat} has the block-diagonal structure
\eqref{eq:block_diag} with both (even and odd) blocks identical
\eqref{eq:block_equal}, we can focus only on the even set $\langle \Phi^e_i |
\mathcal{K}_+^2 | \Phi^e_j \rangle$:
\begin{equation}
\begin{aligned}
\begin{array}{rrrrrrr}
                 &                  &{1}{2}{3}&\bar{1}\bar{2}{3}&\bar{1}{2}\bar{3}&{1}\bar{2}\bar{3}& \\
{1}{2}{3}        &\ldelim({4}{0.5em}&-3    &    2         &   2      &   2      &\rdelim){4}{0.5em} \\
\bar{1}\bar{2}{3}&                  & 2    &   -3         &  -2      &  -2      & \\
\bar{1}{2}\bar{3}&                  & 2    &   -2         &  -3      &  -2      & \\
{1}\bar{2}\bar{3}&                  & 2    &   -2         &  -2      &  -3      & \\
\end{array}
\end{aligned}.
\label{eq:K2+ij3EM}
\end{equation}
Again, the matrix elements are real and the diagonal is equal to minus the
number of unpaired electrons. Diagonalization of this matrix leads directly to
the set of eigenvalues $\{-9,-1,-1,-1\}$ with the corresponding eigenvectors (in
columns)
\begin{equation}
  \left(
    \begin{aligned}
    \begin{array}{rrrr}
     1/2 & \sqrt{3}/2 &     0       &     0       \\
    -1/2 & \sqrt{3}/6 &     0       &  \sqrt{6}/3 \\
    -1/2 & \sqrt{3}/6 &  \sqrt{2}/2 & -\sqrt{6}/6 \\
    -1/2 & \sqrt{3}/6 & -\sqrt{2}/2 & -\sqrt{6}/6 \\
    \end{array}
    \end{aligned}
  \right).
  \label{eq:K2+ij3EevecSym}
\end{equation}
The triply degenerate eigenvectors in \eqref{eq:K2+ij3EevecSym} have been
chosen to mimic as close as possible the nonrelativistic $\mathcal{S}^2$
eigenvectors [see Eq.~\eqref{eq:S2+ij3EevecSym}].  To construct the
eigenvectors for the odd manifold one can apply either of the operators
$\mathcal{K}_+$ and $\mathcal{K}$ [see Sec. \ref{sec:paired} and Eq.
\eqref{eq:even_odd}].  Note that the eigenfunctions are orthonormal, in
contrast to the previously reported eigenfunctions with eigenvalue
$-1$~\cite{Bucinsky2016}.

In Appendix \ref{supplement}, we provide explicit expressions for the case of
four and five open shells and in Supplemental Material~\cite{supplemental2016}
we make available a program for obtaining the appropriate Kramers configuration
state functions for cases of up to 10 unpaired electrons.

%
%
The discussion in Secs. \ref{sec:Kp}--\ref{sec:DC} is valid for systems
described by Dirac-Coulomb and one-component nonrelativistic or scalar
relativistic Hamiltonians. Since spin symmetry is valid in the
one-component domain we can compare the eigenfunctions of both $\mathcal{K}^2_+$
and $\mathcal{S}^2$~\cite{Pauncz1979} operators in more detail. In
one-component theory, the appropriate basis functions are spin-restricted Slater
determinants $\{\Phi^s\}$ in which $\mathcal{S}^2$ has the matrix form
\begin{equation}
  \left( \mathcal{S}^2 \right)_{ij} =
    \big< \Phi^s_i \big| \mathcal{S}^2 \big| \Phi^s_j \big>.
  \label{eq:S2_mat}
\end{equation}
Considering the case of two open-shell electrons $\{\Phi^s_{12},
\Phi^s_{\bar{1}\bar{2}}, \Phi^s_{1\bar{2}}, \Phi^s_{\bar{1}2}\}$ we obtain
\begin{equation}
  \begin{array}{ccccccc}
                 &                  & 12 & \bar{1}\bar{2} & 1\bar{2} & \bar{1}2 & \\
  12             &\ldelim({4}{0.5em}& 2  &    0           &  0       &   0      &\rdelim){4}{0.5em}\\
  \bar{1}\bar{2} &                  & 0  &    2           &  0       &   0      & \\
  1\bar{2}       &                  & 0  &    0           &  1       &   1      & \\
  \bar{1}2       &                  & 0  &    0           &  1       &   1      & \\
  \end{array}
\end{equation}
with eigenvalues and eigenvectors
\begin{equation}
  \def\arraystretch{1.5}
  \begin{array}{ll}
  \Psi^{1, 1} =                            \Phi^s_{12},                                  &\qquad e_1= 2,
  \\
  \Psi^{1,-1} =                                          \Phi^s_{\bar{1}\bar{2}},        &\qquad e_2= 2,
  \\
  \Psi^{1, 0} = \tfrac{1}{\sqrt{2}} \left( \Phi^s_{1\bar{2}} + \Phi^s_{\bar{1}2} \right),&\qquad e_3= 2,
  \\
  \Psi^{0, 0} = \tfrac{1}{\sqrt{2}} \left( \Phi^s_{1\bar{2}} - \Phi^s_{\bar{1}2} \right),&\qquad e_4= 0.
  \end{array}
  \label{eq:spin_func}
\end{equation}
To be consistent with the previous discussion, we have used unbarred ($\alpha$)
and barred ($\beta$) notation for the one-electron spinors. Comparing the eigenfunctions
\eqref{eq:spin_func} and \eqref{eq:K2+ij2eval}, we note that while the
singlet $\Psi^{0, 0}$ and the low-spin triplet $\Psi^{1, 0}$ wave functions
remain unchanged, we need to combine the high-spin triplet wave functions,
$\Psi^{1, 1}$ and $\Psi^{1,-1}$, to obtain the remaining eigenfunctions in
Eq.~\eqref{eq:K2+ij2eval}. Interestingly, this behavior was already observed when
representing one-electron operators in the second quantization formalism.  The
standard excitation operators in the nonrelativistic
theory~\cite{Helgaker-2.2} are
\begin{equation}
  \begin{aligned}
  &\hat T^{1,1}_{pq}  =      - a^\dagger_p a_{\bar{q}},
   \\
  &\hat T^{1,-1}_{pq} =        a^\dagger_{\bar{p}} a_q,
   \\
  &\hat T^{1,0}_{pq}  = \tfrac{1}{\sqrt{2}} \left( a^\dagger_p a_q - a^\dagger_{\bar{p}} a_{\bar{q}} \right),
   \\
  &\hat S^{0,0}_{pq}  = \tfrac{1}{\sqrt{2}} \left( a^\dagger_p a_q + a^\dagger_{\bar{p}} a_{\bar{q}} \right).
  \end{aligned}
  \label{eq:triplet_op}
\end{equation}
When these operators act on the two-electron closed-shell Slater determinant,
spin-adapted wave functions~\eqref{eq:spin_func} are created. On the other hand,
the excitation operators used to describe the one-electron Dirac
operator~\cite{Dyall-chapter9}
\begin{equation}
  \begin{aligned}
  \hat E^-_{\bar{p}q} = \left( a^\dagger_{\bar{p}} a_q + a^\dagger_p a_{\bar{q}} \right),
  \\
  \hat E^+_{\bar{p}q} = \left( a^\dagger_{\bar{p}} a_q - a^\dagger_p a_{\bar{q}} \right),
  \\
  \hat{E}^-_{pq} = \left( a^\dagger_p a_q - a^\dagger_{\bar{p}} a_{\bar{q}} \right),
  \\
  \hat E^+_{pq} = \left( a^\dagger_p a_q + a^\dagger_{\bar{p}} a_{\bar{q}} \right)
  \end{aligned}
\end{equation}
create Kramers configuration state functions \eqref{eq:K2+ij2eval} (up to a
normalization factor). For a definition of the creation and annihilation
operators, see the corresponding literature \cite{Helgaker-2.2, Dyall-chapter9}.
Moreover, the equivalent of the triplet operators $\hat T$ [Eq. \eqref{eq:triplet_op}] in
the nonrelativistic case, known as Cartesian components of triplet excitation
operators~\cite{Helgaker-2.2}, produce wave functions of $\mathcal{K}^2_+$
[Eq. \eqref{eq:K2+ij2eval}]. This reflects the fact that both $\mathcal{K}^2_+$ and
$\mathcal{S}^2$ are appropriate operators for representing the symmetry in the
nonrelativistic theory.

In the case of three open-shell electrons, the matrix representation of
$\mathcal{S}^2$ has a block-diagonal form. Due to the block-diagonal structure, 
it is possible to construct two identical $4\times4$ matrices. In the same basis 
as $\mathcal{K}^2_+$ in Eq. \eqref{eq:K2+ij3EM} one of the matrices
reads as
\begin{equation}
\begin{aligned}
\begin{array}{rrccccc}
                 &                  &{1}{2}{3}&\bar{1}\bar{2}{3}&\bar{1}{2}\bar{3}&{1}\bar{2}\bar{3}& \\
{1}{2}{3}        &\ldelim({4}{0.5em}& 15/4 &   0   &   0   &   0   &\rdelim){4}{0.5em} \\
\bar{1}\bar{2}{3}&                  &   0  &  7/4  &   1   &   1   & \\
\bar{1}{2}\bar{3}&                  &   0  &   1   &  7/4  &   1   & \\
{1}\bar{2}\bar{3}&                  &   0  &   1   &   1   &  7/4  & \\
\end{array}
\end{aligned}.
\end{equation}
The eigenspectrum of this matrix is doubly degenerate
$\{15/4,15/4,3/4,3/4\}$ permitting the freedom of unitary rotation among
the degenerate eigenvectors. It is customary to choose the eigenvectors
corresponding to eigenvalue $15/4$ being simultaneously eigenvectors of
$\mathcal{S}_z$ operator with eigenvalues $3/2$ and $1/2$. The remaining
eigenvectors with $\mathcal{S}_z$ eigenvalue $1/2$ were selected in their conventional 
form~\cite{Pauncz1979} (eigenvectors ordered in columns)
\begin{equation}
  \left(
    \begin{aligned}
    \begin{array}{rrrr}
      1  &      0     &     0       &     0       \\
      0  & 1/\sqrt{3} &     0       &  \sqrt{6}/3 \\
      0  & 1/\sqrt{3} &  \sqrt{2}/2 & -\sqrt{6}/6 \\
      0  & 1/\sqrt{3} & -\sqrt{2}/2 & -\sqrt{6}/6 \\
    \end{array}
    \end{aligned}
    \label{eq:S2+ij3EevecSym}
  \right).
\end{equation}
Similarly to the case of two open-shell electrons, two of the low-spin
eigenvectors $\Psi^{3/4, 1/2}$ are identical to eigenvectors of the
$\mathcal{K}^2_+$ operator [see Eq. \eqref{eq:K2+ij3EevecSym}]. On the other hand,
to obtain remaining eigenfunctions of Eq. \eqref{eq:K2+ij3EevecSym} we need to
combine both low-spin $\Psi^{15/4, 1/2}$ and high-spin $\Psi^{15/4, 3/2}$
eigenvectors, and hence break the $\mathcal{S}_z$ symmetry.


\section{\label{sec:conc} Conclusions}

In this work, we have shown the connection between the recently proposed
time-reversal generator~\cite{Bucinsky2015,Bucinsky2016}
\begin{equation}
  \mathcal{K}_+=\sum_i^N K_i
  \label{eq:Kp_con}
\end{equation}
and the well-known unitary time-reversal operator $\mathcal{K}$
\begin{equation}
  \mathcal{K} = e^{ \frac{\pi}{2} \mathcal{K}_+ }
  \label{eq:KK+}
\end{equation}
for the case of an $N$-electron system.  Based on the relation~\eqref{eq:KK+},
we have proved the eigenvalue theorem for the square of the time-reversal
generator $\mathcal{K}^2_+$ without the need of knowing an explicit form of its
eigenfunctions
\begin{equation}
  \begin{gathered}
  \mathcal{K}_+^2 \Psi = -k^2 \Psi, \qquad k\in\mathbb{N}_0,
  \\
  \begin{array}{ccc}
      \text{odd }  N & \Leftrightarrow & \text{odd } k,
      \\
      \text{even } N & \Leftrightarrow & \text{even } k,
  \end{array}
  \end{gathered}
  \label{eq:K2eigen_con}
\end{equation}
Since $\mathcal{K}^2_+$ operator commutes with the Dirac-Coulomb (DC)
Hamiltonian in the basis of Kramers-restricted Slater determinants
\begin{equation}
  [\bm{H}^\mathrm{DC},\bm{\mathcal{K}}_+^2] = 0,
\end{equation}
the eigenvalues $-k^2$ represent a new quantum number for the relativistic wave
functions and give rise to a new type of symmetry in relativistic many-particle
systems described by DC Hamiltonian
\begin{equation}
  \bm{U} \equiv e^{i \theta \bm{\mathcal{K}}_+^2}.
\end{equation}

Furthermore, the time-reversal generator defines an orthonormal pair of wave
functions $\{\Psi,\widetilde\Psi\}$ which are degenerate eigenfunctions of both
DC Hamiltonian and the $\mathcal{K}^2_+$ operator
\begin{equation}
  \mathcal{K}_+ \Psi \equiv k \widetilde\Psi.
\end{equation}
We have shown the connection between the pair $\{\Psi,\widetilde\Psi\}$ and
standard Kramers pair $\{\Psi,\overbar\Psi\}$. It turns out that while for an
odd-electron system $\overbar\Psi = \pm\widetilde\Psi$, for an even-electron
system $\Psi = \pm\overbar\Psi$ and $\widetilde\Psi \ne \overbar\Psi$.  From
these relations several consequences arise and are related to matrix elements
of operators $\mathcal{O}$ responsible for interactions with magnetic fields
and to the degeneracy of energy levels for the Dirac-Coulomb Hamiltonian. One
especially interesting result holds for systems with an even number of electrons,
where the matrix elements of $\mathcal{O}$ in the basis of
eigenfunctions~\eqref{eq:K2eigen_con} are zero on the diagonal and either pure
real or pure imaginary on the off diagonal.

The general eigenvalue theorem~\eqref{eq:K2eigen_con} was confirmed
analytically in the basis composed of Kramers-restricted Slater determinants
with further restriction on the quantum number $k$ to the number of the
unpaired electrons. A program solving the eigenvalue
problem~\eqref{eq:K2eigen_con} is provided within the Supplemental Material
\cite{supplemental2016}.

The symmetry and corresponding constants of motion presented here, offer a
comparable amount of informations about the relativistic many-electron systems
as the spin quantum numbers in nonrelativistic theory. We therefore believe
that the new quantum number $-k^2$ will prove useful in different areas of
quantum physics. There are several applications we can foresee.
\begin{enumerate}[label=(\roman*)]
\item
Since we have now access to the quantum number $-k^2$~\eqref{eq:K2eigen_con},
it is possible to measure the difference
\begin{equation}
-k^2-\big<\Psi\big|\mathcal{K}^2_+\big|\Psi\big>
\label{eq:Kramer_cont}
\end{equation}
with $\Psi$ obtained from the Kramers-unrestricted solutions of
density-functional theory (DFT) or Hartree--Fock theory (HF).  We call the
measure~\eqref{eq:Kramer_cont} \emph{Kramers contamination}, in analogy to the spin
contamination in nonrelativistic DFT and HF theories, where it is evaluated as
the difference of the $S(S+1)$ spin quantum number and the inner product of
spin-unrestricted wave functions over $\mathcal{S}^2$ operator. The
Kramers contamination has already been studied in the framework of
two-component HF theory in the pilot work which introduced the time-reversal
generator~\eqref{eq:Kp_con}~\cite{Bucinsky2015}.
\item
Characterization of spectra for heavy-element containing compounds and
selection rules based on symmetry generated by the $\mathcal{K}^2_+$ operator.
\item
Kramers configuration state functions (KCSF) as relativistic analogs of the
nonrelativistic configuration state functions (known also as spin-adapted
functions)~\cite{Helgaker-2.5}.
\item
Reduced computational effort associated with the evaluation of operator
matrix elements in the KCSF basis.
\item
Relation between the symmetry generated by the $\mathcal{K}^2_+$ operator and
double-group symmetry.
\end{enumerate}


\section{\label{sec:ackn} Acknowledgments}
Financial support was obtained from VEGA (contract No. 1/0327/12) and APVV
(contract Nos. APVV-15-0079 and APVV-15-0053). This work also received support
from the Research Council of Norway through a Centre of Excellence Grant (Grant
No. 179568) and project grant No. 214095.  Furthermore, the project is financed
from the SASPRO Programme (Contract no.  1563/03/02). The research leading to
these results has received funding from the People Programme (Marie Curie
Actions) European Union’s Seventh Framework Programme under REA Grant Agreement
No. 609427 and has been further cofunded by the Slovak Academy of Sciences.

\appendix

\section{\label{app:0}}

We assume that Kramers-restricted Slater determinants $\{\Phi_i\}$ constitute a
complete basis in the Fock subspace $S^- H^{\otimes N}$.  Thus to define inner
product of two wave functions $\Psi,\Phi \in S^- H^{\otimes N}$, it is
sufficient to define the inner product between two determinants
\begin{align}
  \big< \Phi_i \big| \Phi_j \big> =
  \frac{1}{N!} \sum^{N!}_{\zeta,\xi=1}
  P_L^\zeta P_R^\xi (-1)^{\zeta+\xi}
  \big< \phi_{i1} \big| \phi_{j1} \big>
  \cdots
  \big< \phi_{iN} \big| \phi_{jN} \big>.
\end{align}
Here the permutation operator $P_L$ ($P_R$) acts on the indices of bra (ket)
functions, and the inner product of two one-electron wave functions $\left<
\phi_i | \phi_j \right>$ is defined in Eq. \eqref{eq:1e_inner}.

\section{\label{app:1}}

For clarity, we omit the index $i$ in $K_i$ in the following text.  For a real
number $\theta$ we can then write
\begin{equation}
  \begin{aligned}
    & e^{\theta K} = 1 + \theta K + \frac{\theta^2}{2!}K^2 +
        \frac{\theta^3}{3!}K^3 + \frac{\theta^4}{4!}K^4 +
        \frac{\theta^5}{5!}K^5 + \cdots
    \\
    & = 1 + \theta K + \frac{\theta^2}{2!}(-1) +
        \frac{\theta^3}{3!}(-K) + \frac{\theta^4}{4!} +
        \frac{\theta^5}{5!}K + \cdots
    \\
    & = ( 1 - \frac{\theta^2}{2!} + \frac{\theta^4}{4!} - \cdots) +
        K( \theta - \frac{\theta^3}{3!} + \frac{\theta^5}{5!} - \cdots)
    \\
    & = \cos(\theta) + K \, \sin(\theta),
  \end{aligned}
  \label{eKproof}
\end{equation}
where we have used repeatedly expression~\eqref{eq:Ki2}.

\section{\label{app:2}}

The definition of the adjoint of an antilinear operator in the Fock subspace
$S^- H^{\otimes N}$ reads as
\begin{equation}
  \big< \mathcal{O}^\dagger \Lambda_1 \big|             \Lambda_2 \big> =
  \big<                     \Lambda_1 \big| \mathcal{O} \Lambda_2 \big>^\ast.
  \label{eq:ne_inner}
\end{equation}
Note, however, that Eq.~\eqref{eq:ne_inner} is given in the perspective of
Appendix \ref{app:0}. In that case, the definition of the adjoint of operators $\mathcal{K}$
and $\mathcal{K}_+$ follows from the expression for the adjoint of a
one-electron antilinear operator \eqref{eq:adjoint}. Furthermore, it requires
that the Kramers-restricted determinants constitute a complete basis in $S^-
H^{\otimes N}$ and are constructed from orthonormal four-spinors.

Taking into account the above definition of an adjoint, from Eqs.
\eqref{eq:KvsK+} and \eqref{eq:K2p_def} it follows that
\begin{equation}
  \mathcal{K}_+^\dagger = -\mathcal{K}_+.
\end{equation}
Assuming that the wave function $\Phi$ is a normalized
eigenfunction of $\mathcal{K}_+^2$
\begin{equation}
  \mathcal{K}_+^2 \Phi = \kappa \Phi, \qquad \big< \Phi \big| \Phi \big> = 1,
\end{equation}
and defining the action of the time-reversal generator
\begin{equation}
  \mathcal{K}_+ \Phi = \widetilde\Phi,
\end{equation}
the following statements are straightforward to show
\begin{align}
  \begin{aligned}
    \big< \Phi \big| \widetilde\Phi \big> & =
    \big< \Phi \big| \mathcal{K}_+\Phi \big> =
    \big< \mathcal{K}_+^\dagger\Phi \big| \Phi \big>^\ast
    \\
    &=\big< \Phi \big| \mathcal{K}_+^\dagger\Phi \big> =
    -\big< \Phi \big| \mathcal{K}_+\Phi \big> =
    -\big< \Phi \big| \widetilde\Phi \big>
    \\
    &\Rightarrow \big< \Phi \big| \widetilde\Phi \big> = 0,
  \end{aligned}
\end{align}

\begin{align}
  \begin{aligned}
    \big< \widetilde\Phi \big| \widetilde\Phi \big> &=
    \big< \mathcal{K}_+\Phi \big| \mathcal{K}_+\Phi \big> =
    \big< \mathcal{K}_+^\dagger\mathcal{K}_+\Phi \big| \Phi \big>^\ast
    \\
    &=\big< \Phi \big| \mathcal{K}_+^\dagger\mathcal{K}_+\Phi \big> =
    -\big< \Phi \big| \mathcal{K}^2_+\Phi \big> =
    -\kappa \big< \Phi \big| \Phi \big>
    \\
    &\Rightarrow \big< \widetilde\Phi \big| \widetilde\Phi \big> = -\kappa,
  \end{aligned}
\end{align}

\begin{align}
  \begin{aligned}
    \mathcal{K}^2_+ \Phi &= \mathcal{K}_+\widetilde\Phi = \kappa \Phi,
    \\
    \mathcal{K}^2_+\widetilde\Phi &= \kappa \mathcal{K}_+\Phi
    \\
    &\Rightarrow \mathcal{K}_+^2 \widetilde\Phi = \kappa \widetilde\Phi.
  \end{aligned}
\end{align}

\section{\label{app:3}}

Starting from the connection between the $\mathcal{K}$ and $\mathcal{K}_+$
operators \eqref{eq:Kx_vs_Kp}, using a Taylor expansion and eigenvalue Eq.
\eqref{eq:K2p_eigen1}, we get
\begin{equation}
  \begin{aligned}
  \overbar\Psi
  & = \mathcal{K} \Psi = e^{\frac{\pi}{2} \mathcal{K}_+}\Psi \\
  & = \left[
            1 - \frac{1}{2!} \left( \frac{\pi}{2} k \right)^2
              + \frac{1}{4!} \left( \frac{\pi}{2} k \right)^4 + \cdots
      \right] \Psi \\
  & + \left[
            \frac{\pi}{2} - \frac{1}{3!} \left( \frac{\pi}{2} \right)^3 k^2
                          + \frac{1}{5!} \left( \frac{\pi}{2} \right)^5 k^4 + \cdots
      \right] \mathcal{K}_+ \Psi. \\
  \end{aligned}
  \label{eq:app_3a}
\end{equation}
For $k=0$, the wave function $\Psi$ and its Kramers pair $\overbar\Psi$ are identical:
\begin{equation}
  k=0 \quad \Rightarrow \quad \overbar\Psi = \Psi,
\end{equation}
where we used $\mathcal{K}_+ \Psi = 0$ [see Eq. \eqref{eq:k_zero}].

Considering the definition of $\widetilde\Psi$ [Eq. \eqref{eq:pseudo_pair1}], we can
rewrite expression \eqref{eq:app_3a} for $k\neq0$ as
\begin{equation}
  \begin{aligned}
  \overbar\Psi &= \left[
                                  \cos \left( \frac{\pi}{2} k \right) \hat 1
                    + \frac{1}{k} \sin \left( \frac{\pi}{2} k \right) \mathcal{K}_+
                  \right] \Psi \\
               &= \cos \left( \frac{\pi}{2} k \right) \Psi
                + \sin \left( \frac{\pi}{2} k \right) \widetilde\Psi.
  \end{aligned}
  \label{eq:app_3b}
\end{equation}

We can repeat the same procedure for $\widetilde\Psi$ since it shares the same
eigenvalue $-k^2$ with $\Psi$. After obtaining 
\begin{equation}
  \overbar{\widetilde\Psi} = \cos \left( \frac{\pi}{2} k \right) \widetilde\Psi
                           - \sin \left( \frac{\pi}{2} k \right) \Psi,
\end{equation}
we can combine this expression with Eq. \eqref{eq:app_3b} to get the final result in compact
matrix form
\begin{equation}
  \def\arraystretch{1.3}
  \left(
  \begin{array}{rr}
       \cos\big(\frac{\pi}{2}k\big)  &  \sin\big(\frac{\pi}{2}k\big)  \\
      -\sin\big(\frac{\pi}{2}k\big)  &  \cos\big(\frac{\pi}{2}k\big)
  \end{array}
  \right)
  \left(
  \begin{array}{c}
       \Psi \\
       \widetilde\Psi
  \end{array}
  \right)
  =
  \left(
  \begin{array}{c}
       \overbar\Psi \\
       \overbar{\widetilde\Psi}
  \end{array}
  \right).
  \label{eq:paired_structures_appendix}
\end{equation}

\section{\label{app:4}}

In this appendix, we assume summation over repeated indices, bold symbols
stand for $2\times2$ or $4\times4$ matrices depending on the context,
and the following index notation is employed: $\lambda$, $\tau$,
$\mu$ and $\nu$ denote atomic basis functions and {\it p, q, r, s,} and {\it
t} are molecular orbital functions.

The orthonormal restricted kinetically balanced (RKB)~\cite{Stanton} basis can
be expressed as
\begin{equation}
\bm{X}_\lambda =
  \left(
  \begin{matrix}
      \bm{1}  &  \bm{0}  \\
      \bm{0}  &  \frac{1}{2c} \vec{\bm\sigma}\cdot\vec{p}
  \end{matrix}
  \right) \chi_\tau
  \left(
  \begin{matrix}
      S^{-\frac{1}{2}}_{\tau\lambda} \bm{1} &  \bm{0}  \\
      \bm{0} &  2c \, T^{-\frac{1}{2}}_{\tau\lambda} \bm{1}
  \end{matrix}
  \right),
  \label{eq:RKB}
\end{equation}
where $\chi_\tau$ stands for a Gaussian-type scalar function and
\begin{gather}
  S_{\lambda\tau} = \big< \chi_\lambda \big| \chi_\tau \big>,
  \\
  T_{\lambda\tau} = \big< \chi_\lambda \big| p^2 \big| \chi_\tau \big>.
\end{gather}

Due to to the fact that the RKB basis commutes with the one-electron
time-reversal symmetry (TS) operator
\begin{equation}
  [\bm K,\bm{X}_\lambda] = 0,
  \label{eq:K_RKB}
\end{equation}
the matrix elements of the TS operator have the simple form
\begin{equation}
  \bm{K}_{\lambda\tau} =
  \big< \bm{X}_\lambda \big| K \big| \bm{X}_\tau \big> =
  -i \bm\Sigma_y K_0 \delta_{\lambda\tau}.
  \label{eq:mTS}
\end{equation}

The one-electron Dirac operator \eqref{eq:dirac} in the basis \eqref{eq:RKB}
can be written as
\begin{gather}
  \bm{D}_{\lambda\tau} = \big< \bm{X}_\lambda \big| D \big| \bm{X}_\tau \big>
  \nonumber
  \\
  = \left(
  \begin{matrix}
      \left( c^2 \delta_{\lambda\tau} +
           S^{-\frac{1}{2}}_{\lambda\mu} V_{\mu\nu} S^{-\frac{1}{2}}_{\nu\tau}
      \right) \bm{1}
    &
      c \, S^{-\frac{1}{2}}_{\lambda\mu} T^{ \frac{1}{2}}_{\mu\tau} \bm{1}
    \\
      c \, T^{ \frac{1}{2}}_{\lambda\mu} S^{-\frac{1}{2}}_{\mu\tau} \bm{1}
    &
      -c^2 \delta_{\lambda\tau} \bm{1} +
      T^{-\frac{1}{2}}_{\lambda\mu} \bm{W}_{\mu\nu} T^{-\frac{1}{2}}_{\nu\tau}
  \end{matrix}
  \right),
  \label{eq:mDirac}
\end{gather}
where the external potential matrices are
\begin{gather}
  V_{\lambda\tau} = \big< \chi_\lambda \big| V \big| \chi_\tau \big>,
  \\
  \bm{W}_{\lambda\tau} = \big< \vec{\bm\sigma}\cdot\vec p \,\chi_\lambda
                         \big| V \big|
                               \vec{\bm\sigma}\cdot\vec p \,\chi_\tau \big>.
\end{gather}

That the matrices in Eqs. \eqref{eq:mTS} and \eqref{eq:mDirac} commute is
seen by rewriting the matrix $\bm{W}$ as
\begin{equation}
  \bm{W}_{\lambda\tau} =
    \big< \nabla_l \chi_\lambda \big| V \big| \nabla_l \chi_\tau \big> \bm{1} +
    i \varepsilon_{lmn}
    \big< \nabla_l \chi_\lambda \big| V \big| \nabla_m \chi_\tau \big> \bm{\sigma}_n
\end{equation}
and realizing that
\begin{equation}
  \left[ -i\bm\sigma_y K_0, i\vec{\bm\sigma} \right] = 0.
\end{equation}
We can thus write the commutation relation between the one-electron Dirac
Hamiltonian and the time-reversal operator in the orthonormal RKB basis as
\begin{equation}
  \left[ \bm{K}, \bm{D} \right] = 0.
  \label{eq:DvsK_RKB}
\end{equation}
Since four-spinor molecular orbital coefficients
\begin{equation}
  \varphi_p = \bm{X}_\lambda \bm{C}_{\lambda p}
\end{equation}
act like an unitary transformation from an orthonormal atomic orbital basis
$\bm{X}_\lambda$ to an orthonormal molecular orbital basis $\varphi_p$ we
can write
\begin{equation}
  K_{pq} D_{qr} - D_{pq} K_{qr} = 0.
  \label{eq:1e_comut}
\end{equation}

Thanks to the fact that the Coulomb electron-electron interaction is
represented by a real scalar operator and that the time-reversal operator
commutes with the RKB basis \eqref{eq:K_RKB}, the following identity holds:
\begin{equation}
  K_{pq} g_{qrst} - g_{pqst} K_{qr} + g_{srpq} K_{qt} - K_{pq} g_{srqt} = 0,
  \label{eq:2e_comut}
\end{equation}
where
\begin{equation}
  g_{prst} = \iint r^{-1}_{12} \varphi^\dagger_p(1) \varphi_r(1)
                               \varphi^\dagger_s(2) \varphi_t(2) \, dV_{12}.
\end{equation}

Due to the identities \eqref{eq:1e_comut} and \eqref{eq:2e_comut},
the Dirac-Coulomb Hamiltonian and the time-reversal generator commute in the
Fock subspace $F(M,N)$:
\begin{equation}
  \left[ \hat{H}^\mathrm{DC}, \hat{\mathcal{K}}_+ \right ] = 0,
\end{equation}
where $F(M,N)$ contains all Kramers-restricted Slater determinants obtained by
distributing $N$ electrons among $M$ four-spinors, and the operators
$\hat{H}^\mathrm{DC}$ and $\hat{\mathcal{K}}_+$ have the standard form in the
second quantization formalism~\cite{DyallFaegri2007,Dyall-chapter9}.

Finally, we can write the commutation relations in the basis of the Kramers-restricted
Slater determinants $\{\Phi_i\}$ since these form the complete basis in $F(M,N)$:
\begin{equation}
  \big( \hat{H}^\mathrm{DC}  \big)_{ij}
  \big( \hat{\mathcal{K}}_+ \big)_{jk}
  -
  \big( \hat{\mathcal{K}}_+ \big)_{ij}
  \big( \hat{H}^\mathrm{DC}  \big)_{jk}
  =
  0.
\end{equation}

\section{\label{app:5}}

For Hermitian time-reversal anti-symmetric operators
\begin{gather}
  \mathcal{O}^\dagger = \mathcal{O},
  \\
  \mathcal{K}^\dagger \mathcal{O} \mathcal{K} = - \mathcal{O}
\end{gather}
and wave functions $\Psi$ for which it holds that
\begin{equation}
  \overbar\Psi \equiv \mathcal{K} \Psi = e^{i \omega} \Psi,
\end{equation}
it follows
\begin{align}
  \big< \Psi \big| \mathcal{O} \big| \Psi \big>
  & = \big< \overbar\Psi \big| \mathcal{O} \big| \overbar\Psi \big>
    = \big< \mathcal{K}\Psi \big| \mathcal{O} \big| \mathcal{K}\Psi \big>
  \nonumber
  \\
  & = \big<\Psi\big| \mathcal{K}^\dagger \mathcal{O} \mathcal{K} \big|\Psi\big>^\ast
    = - \big< \Psi \big| \mathcal{O} \big| \Psi \big>^\ast
  \nonumber
  \\
  & = - \big< \Psi \big| \mathcal{O} \big| \Psi \big>
  \\
  & \Rightarrow \quad \big< \Psi \big| \mathcal{O} \big| \Psi \big> = 0.
\end{align}
Note that $e^{i \omega}$ is an arbitrary phase factor and we have assumed that
$\mathcal{O}$ is a linear operator.

\section{\label{app:6}}

To express the operator $\mathcal{K}_+$ in the second quantization formalism,
we start from the general form of the one-particle
operators~\cite{Dyall-chapter9}
\begin{align}
  \mathcal{K}_+ = \sum_{pq}
        &\left[
           (\mathcal{K}_+)_{p           q} \, a^\dagger_p        a_q
         + (\mathcal{K}_+)_{p      \bar q} \, a^\dagger_p        a_{\bar q}
         \right.
        \nonumber
        \\
       &\left.
         + (\mathcal{K}_+)_{\bar p      q} \, a^\dagger_{\bar p} a_q
         + (\mathcal{K}_+)_{\bar p \bar q} \, a^\dagger_{\bar p} a_{\bar q}
        \right].
\end{align}
Utilizing the definition of the barred index $\phi_{\bar p} = K \phi_p$,
realizing that the square of the one-electron time-reversal operator equals
minus one [Eq. \eqref{eq:Ki2}], and that we are working with orthonormal one-particle
functions, we can write
\begin{equation}
  \mathcal{K}_+ = \sum_p \left( a^\dagger_{\bar p} a_p - a^\dagger_p a_{\bar p} \right).
  \label{eq:Kp_secondQ}
\end{equation}
It is now easy to see that the operator in Eq. \eqref{eq:Kp_secondQ} mixes only
determinants with the same number of unpaired electrons and that doubly
occupied Kramers pairs are ignored. Finally, we note that the same observations
holds for the square of the time-reversal generator.

\section{\label{app:7}}

In the case of two open-shell electrons, the action of $\mathcal{K}_+^2$
on the Slater determinant basis reads as
\begin{equation}
\begin{aligned}
&\mathcal{K}_+^2 \Psi_{{1}{2}}=-2\Psi_{{1}{2}}+2\Psi_{\bar{1}\bar{2}}, \\
&\mathcal{K}_+^2 \Psi_{\bar{1}\bar{2}}=2\Psi_{{1}{2}}-2\Psi_{\bar{1}\bar{2}}, \\
&\mathcal{K}_+^2 \Psi_{{1}\bar{2}}=-2\Psi_{{1}\bar{2}}-2\Psi_{\bar{1}{2}}, \\
&\mathcal{K}_+^2 \Psi_{\bar{1}{2}}=-2\Psi_{{1}\bar{2}}-2\Psi_{\bar{1}{2}}.
\end{aligned}
\label{K2+ij2}
\end{equation}

For the case of three open-shell electrons, we have chosen just the even
manifold for our considerations since the odd manifold can be easily
constructed applying the time-reversal operator $\mathcal{K}$ on the following
expressions:
\begin{equation}
\begin{aligned}
&\mathcal{K}_+^2 \Psi_{{1}{2}{3}}=-3\Psi_{{1}{2}{3}}+2\Psi_{\bar{1}\bar{2}{3}}+2\Psi_{\bar{1}{2}\bar{3}}+2\Psi_{{1}\bar{2}\bar{3}},  \\
&\mathcal{K}_+^2 \Psi_{\bar{1}\bar{2}{3}}= 2\Psi_{{1}{2}{3}}-3\Psi_{\bar{1}\bar{2}{3}}-2\Psi_{\bar{1}{2}\bar{3}}-2\Psi_{{1}\bar{2}\bar{3}},  \\
&\mathcal{K}_+^2 \Psi_{\bar{1}{2}\bar{3}}= 2\Psi_{{1}{2}{3}}-2\Psi_{\bar{1}\bar{2}{3}}-3\Psi_{\bar{1}{2}\bar{3}}-2\Psi_{{1}\bar{2}\bar{3}},  \\
&\mathcal{K}_+^2 \Psi_{{1}\bar{2}\bar{3}}= 2\Psi_{{1}{2}{3}}-2\Psi_{\bar{1}\bar{2}{3}}-2\Psi_{\bar{1}{2}\bar{3}}-3\Psi_{{1}\bar{2}\bar{3}}.  \\
\end{aligned}
\label{K2+ij3E}
\end{equation}

\section{\label{supplement}}

In the case of four unpaired electrons, the basis of determinants can be splitted
into two separate independent branches of matrix representation of TR generator squared
\begin{equation}
\{\Phi_{{1}{2}{3}{4}},\Phi_{\bar{1}\bar{2}{3}{4}},\Phi_{\bar{1}{2}\bar{3}{4}},\Phi_{\bar{1}{2}{3}\bar{4}},
   \Phi_{{1}\bar{2}\bar{3}{4}},\Phi_{{1}\bar{2}{3}\bar{4}},\Phi_{{1}{2}\bar{3}\bar{4}},\Phi_{\bar{1}\bar{2}\bar{3}\bar{4}} \}
\end{equation}
and
\begin{equation}
\{\Phi_{\bar{1}{2}{3}{4}},\Phi_{{1}\bar{2}{3}{4}},\Phi_{{1}{2}\bar{3}{4}},\Phi_{{1}{2}{3}\bar{4}},
   \Phi_{\bar{1}\bar{2}\bar{3}{4}},\Phi_{\bar{1}\bar{2}{3}\bar{4}},\Phi_{\bar{1}{2}\bar{3}\bar{4}},\Phi_{{1}\bar{2}\bar{3}\bar{4}} \}.
\end{equation}
The even (odd) basis contains exclusively determinants with even (odd) number of
barred spinors.

For illustration we will consider the action of TR generator squared in the form
\begin{equation}
\mathcal{K}_+^2= -4 \hat{1} + 2 \sum_{i<j} K_i K_j 
\end{equation}
on the following four determinants $\Phi_{{1}{2}{3}{4}}$,
$\Phi_{\bar{1}\bar{2}{3}{4}}$, $\Phi_{\bar{1}{2}{3}{4}}$, and 
$\Phi_{\bar{1}\bar{2}\bar{3}{4}}$:
\begin{equation}
  \begin{aligned}
    \mathcal{K}_+^2 \Phi_{{1}{2}{3}{4}}  = &-4\Phi_{{1}{2}{3}{4}}         + 2\Phi_{\bar{1}\bar{2}{3}{4}}
                                           +2\Phi_{\bar{1}{2}\bar{3}{4}} + 2\Phi_{\bar{1}{2}{3}\bar{4}} \\
                                          &+2\Phi_{{1}\bar{2}\bar{3}{4}} + 2\Phi_{{1}\bar{2}{3}\bar{4}}
                                           +2\Phi_{{1}{2}\bar{3}\bar{4}} + 0\Phi_{\bar{1}\bar{2}\bar{3}\bar{4}}, \\
    \mathcal{K}_+^2 \Phi_{\bar{1}\bar{2}{3}{4}} =&+2\Phi_{{1}{2}{3}{4}}         - 4\Phi_{\bar{1}\bar{2}{3}{4}}
                                                  -2\Phi_{\bar{1}{2}\bar{3}{4}} - 2\Phi_{\bar{1}{2}{3}\bar{4}} \\
                                                 &-2\Phi_{{1}\bar{2}\bar{3}{4}} - 2\Phi_{{1}\bar{2}{3}\bar{4}}
                                                  +0\Phi_{{1}{2}\bar{3}\bar{4}} + 2\Phi_{\bar{1}\bar{2}\bar{3}\bar{4}}, \\
    \mathcal{K}_+^2 \Phi_{\bar{1}{2}{3}{4}} =& -4\Phi_{\bar{1}{2}{3}{4}}         - 2\Phi_{{1}\bar{2}{3}{4}}
                                               -2\Phi_{{1}{2}\bar{3}{4}}         - 2\Phi_{{1}{2}{3}\bar{4}} \\
                                             & +2\Phi_{\bar{1}\bar{2}\bar{3}{4}} + 2\Phi_{\bar{1}\bar{2}{3}\bar{4}}
                                               +2\Phi_{\bar{1}{2}\bar{3}\bar{4}} + 0\Phi_{{1}\bar{2}\bar{3}\bar{4}}, \\
    \mathcal{K}_+^2 \Phi_{\bar{1}\bar{2}\bar{3}{4}} =& +2\Phi_{\bar{1}{2}{3}{4}}         + 2\Phi_{{1}\bar{2}{3}{4}}
                                                       +2\Phi_{{1}{2}\bar{3}{4}}         + 0\Phi_{{1}{2}{3}\bar{4}} \\
                                                     & -4\Phi_{\bar{1}\bar{2}\bar{3}{4}} - 2\Phi_{\bar{1}\bar{2}{3}\bar{4}}
                                                       -2\Phi_{\bar{1}{2}\bar{3}\bar{4}} - 2\Phi_{{1}\bar{2}\bar{3}\bar{4}}.
  \end{aligned}
\end{equation}

Thus, we will obtain the following matrix representation of the $\big< \Phi_x
\big| \mathcal{K}_+^2 \big| \Phi_y \big>$ products for the even basis manifold:
\scriptsize
\begin{equation}
\begin{aligned}
\begin{array}{rrrrrrrrrrr}
                 &                  &
{1}{2}{3}{4}&\bar{1}\bar{2}{3}{4}&\bar{1}{2}\bar{3}{4}&\bar{1}{2}{3}\bar{4}&
{1}\bar{2}\bar{3}{4}&{1}\bar{2}{3}\bar{4}&{1}{2}\bar{3}\bar{4}&\bar{1}\bar{2}\bar{3}\bar{4} & \\
{1}{2}{3}{4}        &\ldelim({8}{0.5em}&-4 & 2 & 2 & 2 & 2 & 2 & 2 & 0 &\rdelim){8}{0.5em} \\
\bar{1}\bar{2}{3}{4}&                  & 2 &-4 &-2 &-2 &-2 &-2 & 0 & 2 & \\
\bar{1}{2}\bar{3}{4}&                  & 2 &-2 &-4 &-2 &-2 & 0 &-2 & 2 & \\
\bar{1}{2}{3}\bar{4}&                  & 2 &-2 &-2 &-4 & 0 &-2 &-2 & 2 & \\
{1}\bar{2}\bar{3}{4}&                  & 2 &-2 &-2 & 0 &-4 &-2 &-2 & 2 & \\
{1}\bar{2}{3}\bar{4}&                  & 2 &-2 & 0 &-2 &-2 &-4 &-2 & 2 & \\
{1}{2}\bar{3}\bar{4}&                  & 2 & 0 &-2 &-2 &-2 &-2 &-4 & 2 & \\
\bar{1}\bar{2}\bar{3}\bar{4}&          & 0 & 2 & 2 & 2 & 2 & 2 & 2 &-4 & \\
\end{array}
\end{aligned}
\end{equation}
\normalsize
with eigenvalues equal to
\begin{equation}
\{-16,-4,-4,-4,-4,0,0,0\}^T 
\end{equation}
\onecolumngrid
and the eigenvector coefficients (ordered in columns)
\scriptsize
\begin{equation}
\begin{aligned}
\begin{array}{rrrrrrrrrr}
\ldelim({8}{0.5em} 
& -0.35355339&  0.60199155&  0.00000000& -0.05524348&  0.36681649&  0.06127536&  0.60927828&  0.00503067 &\rdelim){8}{0.5em} \\
&  0.35355339&  0.36496952&  0.05988493& -0.02197086& -0.60226931& -0.24113491&  0.23364026& -0.51211931 & \\
&  0.35355339& -0.06618721&  0.28381790& -0.64415204&  0.01161054& -0.29172449&  0.23048191&  0.48659522 & \\
&  0.35355339&  0.00476227& -0.64487300& -0.28554088& -0.05081864&  0.59413476&  0.14515612&  0.03055476 & \\
&  0.35355339& -0.00476227&  0.64487300&  0.28554088&  0.05081864&  0.59413476&  0.14515612&  0.03055476 & \\
&  0.35355339&  0.06618721& -0.28381790&  0.64415204& -0.01161054& -0.29172449&  0.23048191&  0.48659522 & \\
&  0.35355339& -0.36496952& -0.05988493&  0.02197086&  0.60226931& -0.24113491&  0.23364026& -0.51211931 & \\
& -0.35355339& -0.60199155&  0.00000000&  0.05524348& -0.36681648&  0.06127536&  0.60927828&  0.00503067 & \\
\end{array}
\end{aligned}.
\end{equation}
\normalsize

The odd basis has the following matrix representation:
\scriptsize
\begin{equation}
\begin{aligned}
\begin{array}{rrrrrrrrrrr}
                 &                  &
\bar{1}{2}{3}{4}&{1}\bar{2}{3}{4}&{1}{2}\bar{3}{4}&{1}{2}{3}\bar{4}&
\bar{1}\bar{2}\bar{3}{4}&\bar{1}\bar{2}{3}\bar{4}&\bar{1}{2}\bar{3}\bar{4}&{1}\bar{2}\bar{3}\bar{4} & \\
\bar{1}{2}{3}{4}&\ldelim({8}{0.5em}&-4 &-2 &-2 &-2 & 2 & 2 & 2 & 0 &\rdelim){8}{0.5em} \\
{1}\bar{2}{3}{4}&                  &-2 &-4 &-2 &-2 & 2 & 2 & 0 & 2 & \\
{1}{2}\bar{3}{4}&                  &-2 &-2 &-4 &-2 & 2 & 0 & 2 & 2 & \\
{1}{2}{3}\bar{4}&                  &-2 &-2 &-2 &-4 & 0 & 2 & 2 & 2 & \\
\bar{1}\bar{2}\bar{3}{4}&          & 2 & 2 & 2 & 0 &-4 &-2 &-2 &-2 & \\
\bar{1}\bar{2}{3}\bar{4}&          & 2 & 2 & 0 & 2 &-2 &-4 &-2 &-2 & \\
\bar{1}{2}\bar{3}\bar{4}&          & 2 & 0 & 2 & 2 &-2 &-2 &-4 &-2 & \\
{1}\bar{2}\bar{3}\bar{4}&          & 0 & 2 & 2 & 2 &-2 &-2 &-2 &-4 & \\
\end{array}
\end{aligned}
\vspace{0.2cm}
\end{equation}
\normalsize
with eigenvalues equal to
\begin{equation}
\{-16,-4,-4,-4,-4,0,0,0\}^T
\end{equation}
and the appropriate coefficients (ordered in columns)
\scriptsize
\begin{equation}
\begin{aligned}
\begin{array}{rrrrrrrrrr}
\ldelim({8}{0.5em} 
&  0.35355339& -0.60199155&  0.00000000&  0.05524348&  0.36681649& -0.06127536& -0.60927828& -0.00503067 &\rdelim){8}{0.5em} \\
&  0.35355339&  0.36496952& -0.05988493& -0.02197086&  0.60226931& -0.24113491&  0.23364026& -0.51211931 & \\
&  0.35355339& -0.06618721& -0.28381790& -0.64415204& -0.01161054& -0.29172449&  0.23048191&  0.48659522 & \\
&  0.35355339&  0.00476227&  0.64487300& -0.28554088&  0.05081864&  0.59413476&  0.14515612&  0.03055476 & \\
& -0.35355339&  0.00476227&  0.64487300& -0.28554088&  0.05081864& -0.59413476& -0.14515612& -0.03055476 & \\
& -0.35355339& -0.06618721& -0.28381790& -0.64415204& -0.01161054&  0.29172449& -0.23048191& -0.48659522 & \\
& -0.35355339&  0.36496952& -0.05988493& -0.02197086&  0.60226931&  0.24113491& -0.23364026&  0.51211931 & \\
& -0.35355339& -0.60199155&  0.00000000&  0.05524348&  0.36681648&  0.06127536&  0.60927828&  0.00503067 & \\
\end{array}
\end{aligned}.
\end{equation}
\normalsize

For the five open-shell electrons case (considering only the even barred basis
manifold) we will obtain the following matrix representation:
\scriptsize
\begin{equation}
\begin{aligned}
\begin{array}{rr rrrr rrrr rrrr rrrr r}
 & & {1}{2}{3}{4}{5}& \bar{1}\bar{2}{3}{4}{5}& \bar{1}{2}\bar{3}{4}{5}& \bar{1}{2}{3}\bar{4}{5}& 
\bar{1}{2}{3}{4}\bar{5}& {1}\bar{2}\bar{3}{4}{5}& {1}\bar{2}{3}\bar{4}{5}& {1}\bar{2}{3}{4}\bar{5}& 
{1}{2}\bar{3}\bar{4}{5}& {1}{2}\bar{3}{4}\bar{5}& {1}{2}{3}\bar{4}\bar{5}& \bar{1}\bar{2}\bar{3}\bar{4}{5}& 
\bar{1}\bar{2}\bar{3}{4}\bar{5}& \bar{1}\bar{2}{3}\bar{4}\bar{5}& \bar{1}{2}\bar{3}\bar{4}\bar{5}& {1}\bar{2}\bar{3}\bar{4}\bar{5}& \\

{1}{2}{3}{4}{5}&\ldelim({16}{0.5em}&-5 & 2 & 2 & 2 & 2 & 2 & 2 & 2 & 2 & 2 & 2 & 0 & 0 & 0 & 0 & 0 & \rdelim){16}{0.5em} \\
\bar{1}\bar{2}{3}{4}{5}&           & 2 &-5 &-2 &-2 &-2 &-2 &-2 &-2 & 0 & 0 & 0 & 2 & 2 & 2 & 0 & 0 & \\
\bar{1}{2}\bar{3}{4}{5}&           & 2 &-2 &-5 &-2 &-2 &-2 & 0 & 0 &-2 &-2 & 0 & 2 & 2 & 0 & 2 & 0 & \\
\bar{1}{2}{3}\bar{4}{5}&           & 2 &-2 &-2 &-5 &-2 & 0 &-2 & 0 &-2 & 0 &-2 & 2 & 0 & 2 & 2 & 0 & \\
\bar{1}{2}{3}{4}\bar{5}&           & 2 &-2 &-2 &-2 &-5 & 0 & 0 &-2 & 0 &-2 &-2 & 0 & 2 & 2 & 2 & 0 & \\  
{1}\bar{2}\bar{3}{4}{5}&           & 2 &-2 &-2 & 0 & 0 &-5 &-2 &-2 &-2 &-2 & 0 & 2 & 2 & 0 & 0 & 2 & \\ 
{1}\bar{2}{3}\bar{4}{5}&           & 2 &-2 & 0 &-2 & 0 &-2 &-5 &-2 &-2 & 0 &-2 & 2 & 0 & 2 & 0 & 2 & \\
{1}\bar{2}{3}{4}\bar{5}&           & 2 &-2 & 0 & 0 &-2 &-2 &-2 &-5 & 0 &-2 &-2 & 0 & 2 & 2 & 0 & 2 & \\
{1}{2}\bar{3}\bar{4}{5}&           & 2 & 0 &-2 &-2 & 0 &-2 &-2 & 0 &-5 &-2 &-2 & 2 & 0 & 0 & 2 & 2 & \\ 
{1}{2}\bar{3}{4}\bar{5}&           & 2 & 0 &-2 & 0 &-2 &-2 & 0 &-2 &-2 &-5 &-2 & 0 & 2 & 0 & 2 & 2 & \\
{1}{2}{3}\bar{4}\bar{5}&           & 2 & 0 & 0 &-2 &-2 & 0 &-2 &-2 &-2 &-2 &-5 & 0 & 0 & 2 & 2 & 2 & \\ 
\bar{1}\bar{2}\bar{3}\bar{4}{5}&   & 0 & 2 & 2 & 2 & 0 & 2 & 2 & 0 & 2 & 0 & 0 &-5 &-2 &-2 &-2 &-2 & \\ 
\bar{1}\bar{2}\bar{3}{4}\bar{5}&   & 0 & 2 & 2 & 0 & 2 & 2 & 0 & 2 & 0 & 2 & 0 &-2 &-5 &-2 &-2 &-2 & \\ 
\bar{1}\bar{2}{3}\bar{4}\bar{5}&   & 0 & 2 & 0 & 2 & 2 & 0 & 2 & 2 & 0 & 0 & 2 &-2 &-2 &-5 &-2 &-2 & \\ 
\bar{1}{2}\bar{3}\bar{4}\bar{5}&   & 0 & 0 & 2 & 2 & 2 & 0 & 0 & 0 & 2 & 2 & 2 &-2 &-2 &-2 &-5 &-2 & \\ 
{1}\bar{2}\bar{3}\bar{4}\bar{5}&   & 0 & 0 & 0 & 0 & 0 & 2 & 2 & 2 & 2 & 2 & 2 &-2 &-2 &-2 &-2 &-5 & \\ 
\end{array}
\end{aligned}
\label{eq:5el}
\end{equation}
\normalsize
with eigenvalues being equal to
\begin{equation}
\{-25, 5 \times (-9), 10 \times (-1)\}^T.
\end{equation}

The matrix representation of $\mathcal{K}_+^2$ in the odd basis manifold
\eqref{eq:even_odd} of the five open-shell case is identical to
Eq.~\eqref{eq:5el} [see Eq.~\eqref{eq:block_equal}]. The eigenvectors of the
five open shell case are not shown for brevity. The interested reader might use
the attached \texttt{FORTRAN} code~\cite{supplemental2016} to obtain
eigenfunctions and eigenvalues of $\mathcal{K}_+^2$ for cases with up to 10
unpaired electrons.

\twocolumngrid

\hspace{1cm}

\providecommand{\noopsort}[1]{}\providecommand{\singleletter}[1]{#1}%

\end{document}